# ASSEMBLING THE FRAGMENT D ON THE ANTIKYTHERA MECHANISM: ITS ROLE AND OPERATION IN THE DRACONIC GEARING

**Aristeidis Voulgaris*[1], Christophoros Mouratidis[2] and Andreas Vossinakis[3]**

[1]*City of Thessaloniki, Directorate Culture and Tourism, Thessaloniki, GR-54625, Greece*
[2]*Merchant Marine Academy of Syros, Ermoupoli, GR-84100, Greece*
[3]*Thessaloniki Astronomy Club, Thessaloniki, GR-54646, Greece*

**ABSTRACT**

The unplaced Fragment D of the Antikythera Mechanism with an unknown operation was a mystery ever since it was discovered. The gear r1, which was detected in Fragment-D radiographies by C. Karakalos is preserved in excellent condition, but this was not enough to directly relate it to the existing gear trains of the Mechanism. The suggestion that Fragment D could be a part of the hypothetical planet indication gearing is a hypothesis, in contradiction with the AMRP tomographies, and leads to mechanical malfunctions. The analysis of AMRP tomographies of Fragment D and its mechanical characteristics, revealed that it could be part of the Draconic gearing. Currently, the three lunar cycles Synodic, Sidereal and Anomalistic are represented on the Antikythera Mechanism. Although the Draconic cycle was well known during the Mechanism's era as representing the critical fourth Lunar cycle, it seems that it is missing from the Antikythera Mechanism. The study of Fragment D was supported by our bronze reconstructions of the Fragment D, leading to the reconstruction of the Draconic gearing. These results are very important, because with the use of only one hypothetical/missing gear, the Draconic gearing is revealed on the Antikythera Mechanism. Fragment D(*raconic*) can be perfectly correlated with the Draconic cycle of the Antikythera Mechanism without mechanical problems and the four lunar cycles appear on the Mechanism. Based on the Draconic pointer, the eclipse events can be calculated and engraved on the Saros spiral cells.

**KEYWORDS:** Fragment D, Lunar cycles, Draconic gearing, Nodes, eclipse prediction, Antikythera Mechanism gearing

## 1. INTRODUCTION

The Antikythera Mechanism, a creation of an ingenious manufacturer of the Hellenistic era, was a geared machine capable of performing complex astronomical calculations, mostly based on the lunar periodic cycles Seiradakis and Edmunds (2018). After 2000 years under the Aegean/Ionian sea, Jones (2017); Voulgaris et al., (2019b), the fragments of the Antikythera Mechanism is now a permanent exhibit at the National Archaeological Museum of Athens, Greece.

At ancient time, the Synodic lunar cycle was used as the primary time unit of calendars Hannah (2013). Each month of the ancient Greek calendar was based on the New Moon and Full Moon phases Bowen and Goldstein (1988); Hannah (2013). Even the Olympic Games were occurred every 4 years, but started on the 8th (or 9th) Full Moon after the Winter Solstice, every 49 or 50 lunar synodic months Vaughan (2002). Although one tropical year does not include an integer number of lunar Synodic months, the lunar cycle prevailed instead of the tropical year, which was considered less important Theodosiou and Danezis (1995).

The Metonic calendar known as *Enneadecaeteris* or *Enneakaidekateris* (from Ancient Greek: Εννέακαιδεκαετηρίς: 9+10 έτη-years) correlates the integer number of 235 lunar Synodic cycles (also 254 Sidereal cycles), to the integer number of 19 tropical years. Each Metonic year of 12 or 13 Synodic months deviated from the actual start of the tropical year by several days, in order to keep the lunar Synodic month intact Manitius (1880); van der Waerden (1984b); Theodosiou and Danezis (1995); Spandagos (2002); Freeth et al., (2008); Anastasiou et al., 2016b.

The ancient astronomers chose the lunar cycles instead of the solar cycles for several obvious reasons: it is easy to calculate a time span based on the Moon's every day changing phase, it is visible both at night and at day and when the Moon is on the sky it is easy to observe bright stars. In contrast, the Sun does not exhibit any phases, it is impossible to detect stars when the Sun is above the horizon and the relative position of the Sun on the sky changes by about 1/12th of the lunar angular velocity. Also a particular "solar phase" i.e. a solar eclipse is a rare astronomical event to observe from a specific place. Thus, the ancient astronomers chose to base their measurements and time keeping events on the easier to follow lunar Synodic cycle.

By studying the Antikythera Mechanism it is obvious that the operation, design and the gear teeth selection, were mostly based on the Synodic lunar month: the $b_{in}$ axis-Lunar Disc-Input (the fastest of the rotating axes), the small Lunar phases sphere on the Lunar Disc and the cells on the two spirals (Saros and Metonic), were all precisely based on the Synodic lunar month. The selection of the teeth number of most of the gear trains was made in accordance to the Synodic and Sidereal lunar cycles, e.g. the number of the gear teeth e3 (223 Synodic months of Saros) and ((c2/d1)*d2 = (2 X 127) = 254 Sidereal months of the Metonic cycle). The Lunar Disc rotation, which is the proper driving/input for the Antikythera Mechanism gearing trains (see *Section 4.1*), essentially defines the two lunar cycles - the Sidereal and Synodic month - as the basic time units for the calculations of the Mechanism. Today, only one out of the seven pointers of the Mechanism is clearly related to the tropical year: the Golden Sphere-Sun pointer, Freeth et al., (2006); Voulgaris et al., (2018b); Voulgaris et al., (2019b), i.e. one full rotation equals one tropical year.

All of the above lead to the conclusion that the Antikythera Mechanism was a Luni-(solar) time/calendar geared machine computer, based on the lunar Synodic cycle.

## 2. RESEARCH AIM

The authors' research aims to the detection of a proper position for the unplaced Fragment D/gear r1, on the existing gearing trains of the Mechanism. Today, the preserved/necessary gears of the Mechanism are 35 (+1 unplaced gear/Fragment D). Freeth and Jones (2012) and Freeth et al., (2021), suggested that the Fragment D/gear r1, could be a part of the hypothetical planet indication gearing, representing the planet Venus motion on the Antikythera Mechanism. Freeth et al., (2021) suggested 3D representation design of Fragment D, differs from the AMRP tomographies and presents critical mechanical problems when we constructed it in bronze (see *Section 5.2* and in authors' **Supplementary Material** at the end). For the hypothetical planet gearing design, a large number of hypothetical gears (>33) and hypothetical additional moving parts (>40), is needed. However, today not one of these is preserved. The suggested planet indication gearing on the Antikythera Mechanism is still a hypothesis, requiring a lot of assumptions and presents critical mechanical problems concerning the functionality of the device (see *Section 4.1*).

In this work we will present that the hypothesis of the planet indication gearing is not the only-one, in order to be justified the gear r1 operation and the Fragment D cannot be considered as a proof for the planet indication gearing (see *Section 5.2*) [see also, **Supplementary Material** at the end].

Our present research aim is based on specific data were related to:

a) The knowledge of the Hellenistic Astronomy of the Antikythera Mechanism era (around 200BC-100BC),

b) The mechanical characteristics, properties, specifications and mechanical limitations of the Antikythera Mechanism,

c) The specific dimension of the Mechanism parts,

d) The existence of the real preserved-unplaced parts of the Mechanism.

Finally, our results answer to the question "*how did the ancient Manufacturer calculate the eclipse events and the hours on which they occurred, presented on the Saros spiral cells of the Antikythera Mechanism?*" applying the minimum of possible hypotheses and adapting the minimum of possible hypothetical parts.

## 3. LUNAR CYCLES AND ANTIKYTHERA MECHANISM

### 3.1. *The lunar motions studied in the Hellenistic era*

The four well known lunar motions - the Synodic, Sidereal, Anomalistic and Draconic cycles - where extensively observed and studied by the ancient astronomers, with the aim to interrelate them Spandagos (2002); Steele (2000a,b); Hannah (2001); Steele (2002). *Ptolemy* in *Almagest* Toomer (1984) extensively referred to the primary lunar cycles, giving a large number of calculations in tables, Pedersen (2011). The attempt of *Ptolemy*, *Hipparchus* and the previous era astronomers to (better) correlate - incorporate the four lunar cycles, was obvious.

The ancient astronomers realized that a interrelation - phase synchronization (periodicity) of two of the four lunar motions/cycles, revealed a repeatability of the solar/lunar eclipses, while the synchronization of three lunar cycles, presented eclipses with highly similar geometrical characteristics and classification Oppolzer (1962); van den Bergh (1955); Meeus et al., (1966); Neugebauer (1975); Meeus (1998, 2004). The periodicity of the lunar motions, led to the adoption of the "*interrelated cycles ratio in integer numbers*", a very useful canon, which was the key for the solar/lunar eclipses prediction. *Ptolemy* refers to the Saros cycle with the name "*Περιοδικός*" (*Periodic*), Voulgaris et al., (2021), since it is the smallest single period which contains an integer number of returns of the various motions.

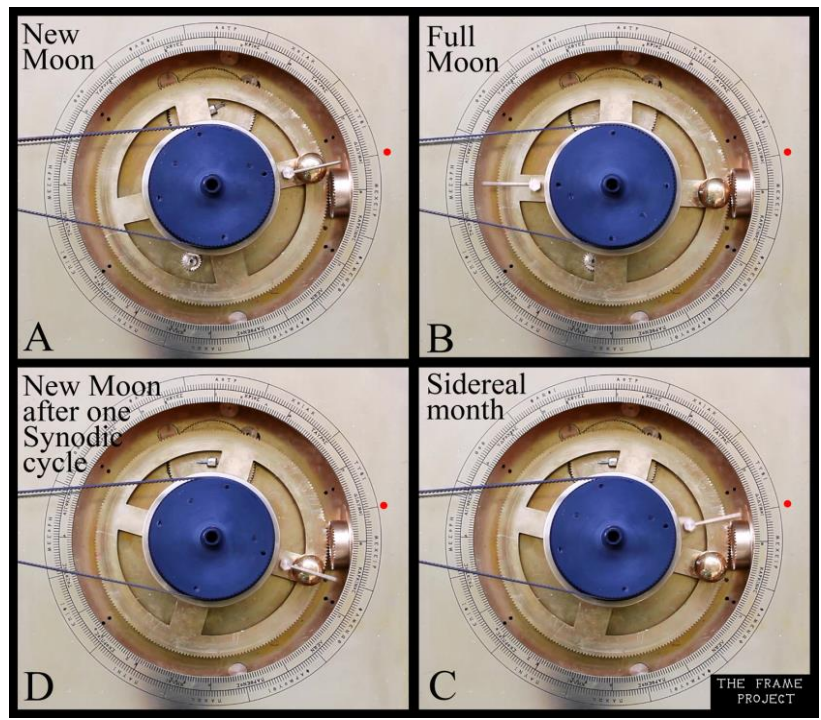

*Figure 1. For our research needs, a pulley is adapted on the Lunar Disc of the Antikythera Mechanism functional model (designed/constructed by the FRAMe Project) and an electrical motor rotates the Lunar Disc via a belt (images taken from video frames). The position/phases of the Moon relative to the Golden sphere-Sun and the zodiac sky are presented. A) The red dot that marks the Egyptian month of 1st MECHIR (ΜΕΧΕΙΡ) and also the 18th zodiac day of Gemini (ΔΙΔΥΜΟΙ) is the starting position of the Lunar Disc and the Golden sphere (as the Egyptian and the Zodiac dial ring are freely rotated, we can turn them to any direction, for our research needs, Voulgaris et al., (2018a). B, C, D) During the rotation of the Lunar Disc, the Golden sphere continuously changes its position relative to the zodiac sky (the slight mismatch of the aiming is a cause of the parallax error during the video recording). Each Synodic month on the Antikythera Mechanism, started right after the New Moon phase. We operate our functional models of the Antikythera Mechanism, by rotating the Lunar Disc. The Lunar Disc offers a slow and precise motion of the pointers, and has a high torque in order to overcome the gearing frictions, as opposed to the hypothetical consideration of the crank-a1 being the input of the Mechanism, see Section 4.1. and Voulgaris et al., (2018b).*

Certainly, the most easily observed lunar cycle was the Synodic month i.e. the period between two successive New Moons or Full Moons, which was easily measured with sufficient accuracy Fig. 1A,D. For this reason the calculations of the rest three lunar cycles were based on the Synodic month.

The Sidereal month is the time duration between two successive transits of the Moon by the same constellation area/star Fig. 1A,C, but the start of each Sidereal cycle presents a different lunar phase. As the Moon is the brighter celestial body in the sky, it far outweighs the bright stars.

The Anomalistic month became evident by the variability of the angular velocity of the Moon against the fixed stars. *Geminus* mentions that at the start of the Anomalistic month (*Apogee*), the lunar angular velocity is 11° 06′ 35″/1d (minimum), whereas at the middle of the Anomalistic month (*Perigee*) the angular velocity is 15° 14′ 35″/1d (maximum). As it may be expected, this angular variation is not easily detected without the use of an astronomical instrument (astrolabe) and the recording of a number of observations and measurements, as *Ptolemy* does in *Almagest*. *Geminus* not only describes the mathematical process for calculating the mean Lunar angular velocity, but also refers to the measuring error as a result of the instrument's limited resolution in practice.

*Table 1. The time measuring cycles expressed in days and in tropical years, as recorded by the observations of Babylonians and the ancient Greek astronomers. True and rounded values of the Draconic month are presented. The 1/1000th of a Synodic month is about 42.5 minutes. The ratio of Draconic to Synodic month is rounded to 1.08520.*

| Cycle (invented for…) | Tropical years | Number of Days | Days per Year | Synodic months | Draconic months (rounded) | Ratio (Draconic/ Synodic) | Draconic month (days) |
|---|---|---|---|---|---|---|---|
| **Saros** (eclipses) | 18y 11.3d | 6585.322 | 365.2233 | 223 | 242 | 1.0852017 | 27.212074 |
| | | | | | 241.999 (true) | 1.0851973 | 27.212186 |
| **Metonic** (tropical year) | 19y | 6940 | 365.2631 | 235 | 255 | 1.0851063 | 27.215686 |
| | | | | | 255.021 (true) | 1.0851957 | 27.213445 |
| **Exeligmos** (eclipses) | 54y 33d | 19756 | 365.2235 | 669 | 726 | 1.0852017 | 27.212121 |
| | | | | | 725.996 (true) | 1.0851958 | 27.212271 |
| **Callippic** (tropical year) | 76y | 27759 | 365.25 | 940 | 1020 | 1.0851063 | 27.214705 |
| | | | | | 1020.84 (true) | 1.0851957 | 27.192312 |
| **Hipparchic** (tropical year) | 345y | 126010 | 365.246 | 4267 | 4630.5 | 1.0851886 | 27.213043 |
| | | | | | 4630.531 (true) | 1.0851959 | 27.212861 |
| **Babylonian** (eclipses) | 441y +106.3d | 161188 | 365.2646 | 5458 | 5923 | 1.0851960 | 27.213911 |
| | | | | | 5922.999 (true) | 1.0851958 | 27.213916 |

The variable lunar motion is also included on the Antikythera Mechanism, introduced by the operation of the *pin&slot* design on the gears k1/k2, off-axis on board the gear e3, Freeth et al., (2006); Wright (2006); Gourtsoyannis (2010); Voulgaris et al., (2018b). The centers of gears k1 and k2 are located at an "off-axis" position and the movement transmission from gear k2 (slot) to k1 (pin) presents a variable angular velocity representing the Anomalistic lunar cycle.

The fourth, more complex lunar motion is the time period that the Moon crosses the Ecliptic. The elliptical lunar orbit is inclined relative to the Ecliptic by about 5.15° and crosses the Ecliptic plane at two points, named *Ascending* and *Descending Node*. *Ptolemy* in *Almagest* calls the time span in which the Moon crosses the same Node, as "Ἀποκατάστασις κατὰ πλάτος" (return to the same latitude), better known to us Draconic or Draconitic month (from Middle Ages myth of the Dragon which "eats" the Sun during a solar eclipse, see Kircher 1646, page 548), also known as nodal or nodical month. In Greek language is used the word Draconic (Δρακωνικός, instead of Draconitic, Δρακωνιτικός), not in compliance to the word Anomalistic (Ἀνωμαλιστικός). The duration of the Draconic month is about 27.2122d, a bit shorter than the Sidereal month (27.3218d) Barbieri (2017). The two Nodes return to the same position relative to the stars after 18.612 tropical years (larger than a Saros cycle, see Espenak and Meeus, 2008-NASA eclipse page). This implies that the Nodes change their projected position on the sky by about 1.564°/1 Synodic month (but not at constant velocity, due to the Anomalistic cycle), transiting each constellation in about 19.2 Synodic months.

As the lunar angular velocity varies, the Moon comes to conjunction, opposition and the Nodes at different velocity each time.

The Metonic, Callippic and Hipparchic cycles were invented in order to better correlate the Synodic month to the tropical year Oppolzer (1962); Meeus et al., (1966); Neugebauer (1975); Meeus (2004). Saros and Exeligmos both fit integer numbers of Draconic and Synodic cycles, but not an integer number of tropical years, and for this reason Saros is the proper choice for the eclipse predictions.

### 3.2. *Requirements for a solar/lunar eclipse - The Draconic cycle*

One of the most important astronomical events in antiquity as well as today is a solar eclipse (https://sites.williams.edu/eclipse/archive/, Voulgaris et al., 2022). In ancient Mesopotamia, Egypt and Greece, eclipses were correlated to an extended mythology regarding the Gods' fight, the Kings, their thrones etc. As Herodotus mentions, the war between Medes and Lydians stopped because of the eclipse of 28 May 585 BC, which Thales had predicted, Panchenko (1994); Stephenson and Fatoohi (1997). The prediction of a solar eclipse was a considerable challenge for the astronomers of the ancient world, who used their observations and calculations to improve the accuracy of the eclipse predictions, Steele (2015).

A solar/lunar eclipse will occur only if two specific astronomical positional parameters are satisfied:

1) The Moon is at its new phase (or Full Moon) and
2) At the same time, the New Moon (or Full Moon) is located at or close to the Ascending or the Descending Node, i.e. close to the beginning or the middle of the Draconic month.

The resonance of these two periods in (about) 0 or π phase, guarantees that a solar (or lunar) eclipse will happen somewhere on the Earth. Of course, in order to predict where on Earth the eclipse will be visible, additional calculations and parameters are needed.

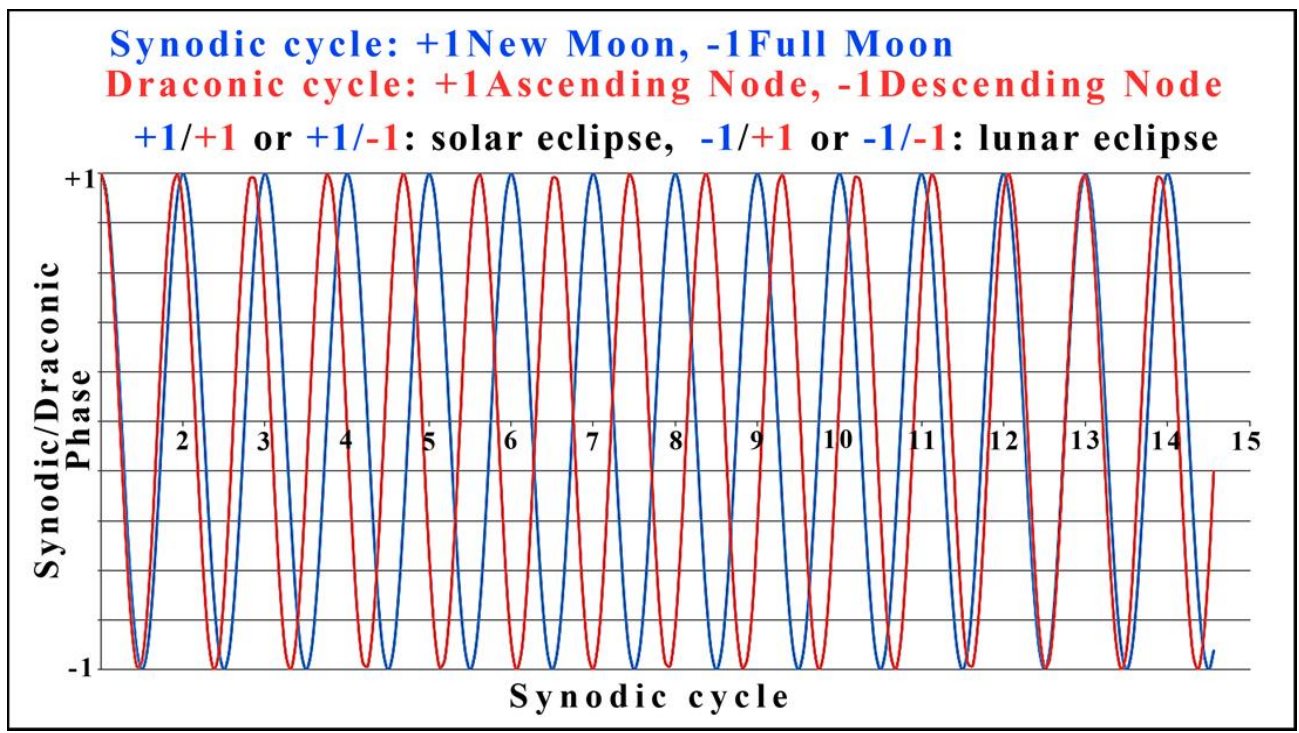

*Figure 2. Harmonic graph of the Draconic vs Synodic cycle, assuming that the two cycles start simultaneously. On each of the resonance (or close to resonance) points (phase π or 2π) of the two graphs, a solar or lunar eclipse will occur. The units on the x-axis are Synodic lunar cycles. (On the calculations for the graph presentation, the phase variation resulting from the Anomalistic cycle was not included).*

As the Moon and the Earth have angular dimensions (about 0.5° for the Moon as observed from the Earth and about 1.5° for the Earth as observed from the Moon), their corresponding shadows are extended. Hence, an eclipse can occur even if the Moon is located at some angular distance from the Node. Thus, a zone on the Ecliptic can be defined with angular dimensions of about ±17° 25′ on either side of the Nodes (for all locations on Earth), called the Zone of Eclipses or Ecliptic limits. The ecliptic limits vary depending on the angular dimensions of the Moon, as a result of its varying distance from the Earth (Apogee/Perigee) i.e. by the Anomalistic cycle Fig. 2.

The exact phase of the Anomalistic cycle (Apogee/Perigee position) during the resonance of the Synodic and Draconic cycles, defines the type of the eclipse event (total, annular or hybrid, also less affected by the Earth's Aphelion/Perihelion), Espenak and Meeus (2008) (NASA eclipse page) and the exact time in which the Moon approaches the Nodes or crosses the ecliptic limits, acting as a variable timer (see *Section 5.4*).

The lower half area of the Back plate of the Antikythera Mechanism was dedicated to the eclipse events information, Freeth et al., (2008); Anastasiou et al., (2016b); Freeth (2019); Iversen and Jones (2019), presented on the Saros four turns spiral, divided into 223 cells/Synodic months. The ancient Manufacturer calculated, designed and constructed the proper gear train for this operation.

At the same time, the Manufacturer engraved the specific sequence of eclipse events on the Saros spiral,
   i) By copying the eclipse information written on a papyrus (?),
   ii) By using the Babylonian eclipse records (?) Carman and Evans (2014); Iversen and Jones (2019)
   or
   iii) The eclipse events were directly (self)calculated making use of his own construction (?) (see *Section 7*).

### *3.3. Eclipses information/prediction mechanism on the Antikythera Mechanism*

The Antikythera Mechanism could predict, the phases of the Moon, the position of the Sun on the sky as it is projected onto a Zodiac constellation, at the corresponding Egyptian and Zodiac month/date, Freeth et al., (2006); Voulgaris et al., (2018a), and also the Metonic month. Additionally it could predict the year and the date in which the athletic games would occur, Freeth et al., (2008).

By turning the Lunar Disc, which is the ideal input for handling the Mechanism Voulgaris et al., (2018b), see *Section 4.1*), the user of the Mechanism could set the lunar pointer aiming directly to the Golden sphere (or to opposite position). At these positions the small half white/half blackened lunar phase sphere of the Lunar Disc shows its black (white) hemisphere, Wright (2006); Carman and DiCocco (2016); Voulgaris et al., (2018b).

*Geminus* in Chap. "*VIII-About months*" and "*XI-About the Lunar eclipse*" mentions that the Full Moon occurred in mid-month, *Διχόμηνις*, and in Chap. "*IX-About the light of the Moon*", and "*X- About the solar eclipse*", that the New Moon occurred at the last day of the Synodic month, on (29th) or 30th day, named *Τριακάς* (Triakas), Theodosiou and Danezis (1995); Spandagos (2002); Jones (2017). Therefore, in each Synodic month, the New Moon phase comes after the Full Moon.

By each re-aiming of the Lunar Disc pointer to the Golden sphere (or to the opposite position-Full Moon), there was a possibility that a solar (or lunar) eclipse would occur. Just by observing the Front dial plate of the Mechanism it was not possible to know for sure if a solar or a lunar eclipse would occur. Information but, not calculation about upcoming eclipses was only presented on the Saros spiral, on the Back plate of the Mechanism, Freeth et al., (2008); Anastasiou et al., (2016b); Iversen Jones (2019) and New Saros cell numbering by Voulgaris et al., (2021). It seems that the calculation of the Draconic cycle - the fourth lunar cycle (well known in Ancient Greece, Egypt and Babylon), which is quite critical for the eclipses prediction - is not presented or preserved on the Antikythera Mechanism.

### *3.4. Two speeds on the Antikythera Mechanism machine gearing*

In the following Section the "nature" of the Antikythera Mechanism as a measuring instrument is to be presented.

By rotating the Lunar Disc-Input of the Mechanism about 389°, one Synodic month is completed Fig. 1. In addition, Saros pointer crosses one cell out of 223 (and the Metonic pointer one out of 235 cells), Anastasiou et al., (2014). The pointer's angular shift is about 6.4°/cell-Synodic month Fig. 3. Thus, the same time unit – a synodic month – is presented by different pointers on the Mechanism.

The Front plate pointers move relatively fast, making an extensive route, offering High- Resolution information regarding their geometrical position. In contrast, the Back plate pointers are Low-speed with Low-positional resolution.

A large number of compressed information is also engraved on the scales of the Low-Speed pointers: 235 words (months) on 235 Metonic spiral cells and a number of eclipse events on some of the 223 Saros spiral cells.

Each Saros (and Metonic) cell corresponds to 29.53d (mean duration). The mean dimensions of these cells are about 7mm×4mm. This means that the Saros pointer changes its position about 6.4°/29.53d ~ 0.257°/1d, Fig. 3. It is impossible to detect/measure time with precision of a day directly on each of the Metonic/Saros cells (7mm/29.53 ~ 0.216mm/1d).

If the pointer aims at the middle of the cell, this does not mean that the date is precisely at the middle of the month. This calculation "inconsistency" arises from the short dimension of each minimum unit, the engagement of the gears with triangular teeth shape, the gear periodic errors, the small precession of the axes, the (slight) off-center positioning of the central gear holes, the random teeth shape mismatches, the mechanical limits of the specific gear train etc., Edmunds (2011). Therefore, the Saros and Metonic spirals' minimum measurable time unit is one cell per one Synodic month.

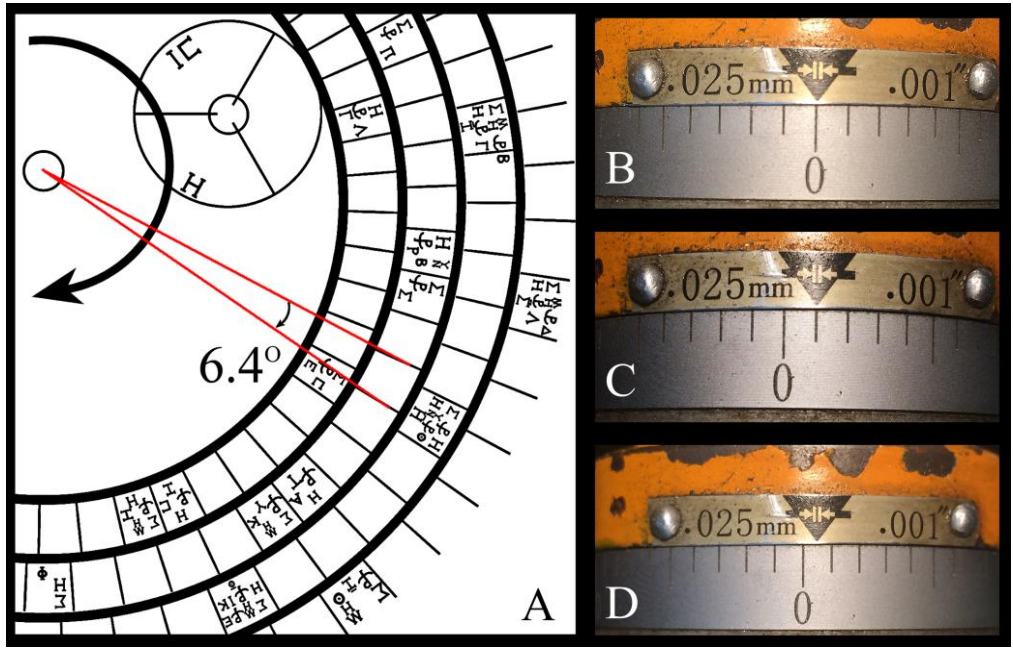

*Figure 3. A) A close up of the Saros spiral with the eclipse information. The angle which the Saros pointer travels in one cell/Synodic month is about 6.4°. Note also the two pairs of successive cells in which the solar eclipse (H) is mentioned on the preceding cell and the lunar eclipse (Σ), to the following cell. Also, the cells with two eclipse events on the same Synodic month are visible (the lunar eclipse on the first line and the solar on the second line). B) Close up of the micrometer dial of the tool post slide hand-wheel of a conventional lathe. C) Each line of the micrometer dial changes the tool post slide position by 0.025mm. D) The position between two lines, the movement of the tool post slide could not be defined as 0.025mm/2, because of the mechanical limitation of the system, this movement presents an uncertainty.*

Finally, the speed of a gear train output defines the resolution of the measurement. The Low-Speed pointers manage an extensive information capacity but in a low geometrical/positional resolution. Therefore, the Saros spiral is more of an "eclipse information table", rather than an eclipse prediction scale. Likewise the Metonic spiral is a month name information table.

It seems that the Antikythera Mechanism lacks High-speed gear train/high-resolution calculations for eclipse predictions that are exclusively based on the geometry of the position.

The authors believe that the ancient Manufacturer would not rely on a pointer of "eclipse information table" to predict the eclipses - the most important events of the Mechanism - without taking into account geometrical/high-resolution calculations for these events.

The existence of two different speeds on the Mechanism leads the authors to consider that the Antikythera Mechanism could have had an additional gear train, with a pointer dedicated to the eclipse prediction calculation. This pointer should be the output of a High-speed gearing train, based on geometry, offering a high-resolution calculation so that the eclipse prediction results would be highly accurate.

### 3.5. *The Fragment D - description and analysis of the gear r1*

Fragment D is an enigmatic and unplaced part of the Antikythera Mechanism. It was first noted by I. Svoronos and photographed by A. Rehm in 1905/1906. It was then misplaced and it was re-found in 1973, Price (1974); Lazos (1994); Freeth and Jones (2012). Fragment D is covered by a layer of calcified deposits, visible by the naked eye. Its mechanical design information cannot be clearly detected by the naked eye Fig. 4.

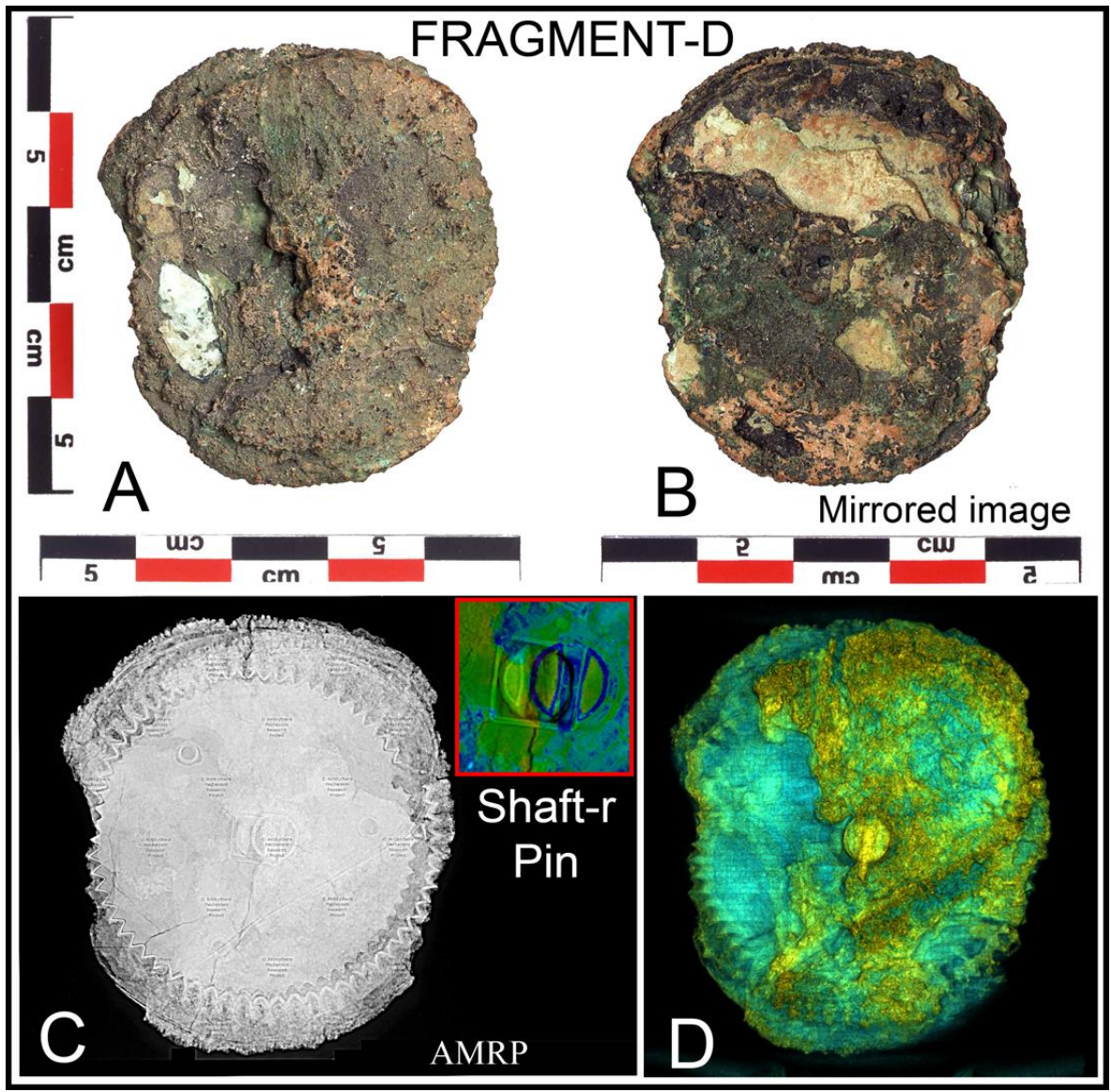

*Figure 4. A) The front (original image) and B) back (mirrored image) visual photographs of Fragment D (Credits: National Archaeological Museum, Athens, K. Xenikakis, Copyright Hellenic Ministry of Culture & Sports/Archaeological Receipts Fund). C) Same scale AMRP corresponding radiography (enhanced). Insert: the displacement of the upper part of the shaft-r and its stabilizing pin is evident (combined tomographies in pseudocolors). D) 3D reconstruction in pseudocolors using AMRP tomographies via 3D Slicer software, Fedorov et al., (2012). The shape of the upper plate is visible. Tomographies processed by the authors.*

AMRP Tomographies revealed that Fragment D consists of three parts, Freeth and Jones (2012). The first part is a partially preserved Circular plate, with a diameter of 43mm (thickness about 1-1.5mm), and a square central hole. On this Circular plate, three perpendicular pins were placed symmetrically at epicenter angles of 120° and are clearly detected Fig. 5. By observing the preserved circular perimeter shape, one concludes that this plate could not be a gear or a scale, as there is no detection of any tooth or engraved line or subdivision(s) on the plate's surface.

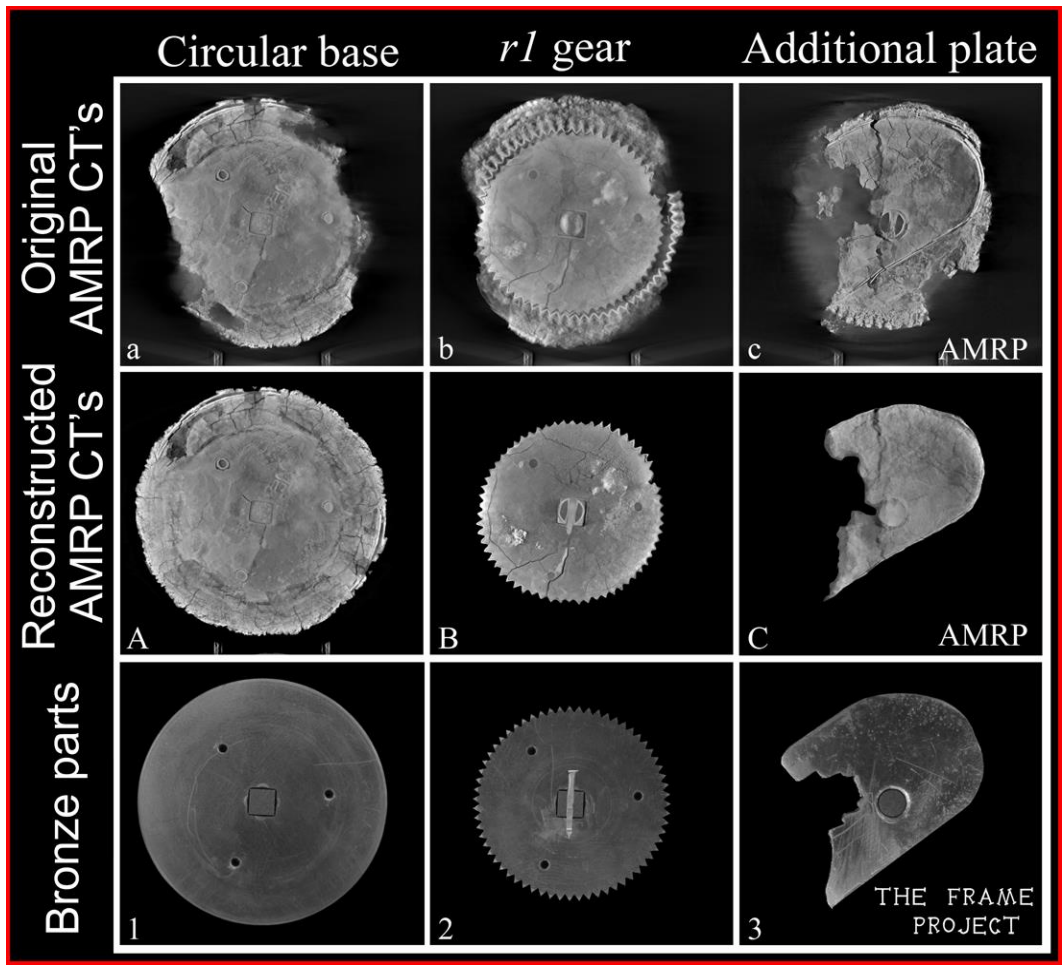

*Figure 5. Selected AMRP tomographies of a) the Circular plate, b) the gear r1, c) the Additional plate. A) the digital reconstruction of the Circular plate tomography, using well-preserved areas of the plate, B) the digital "cleaning" of r1 gear, from its corrosion and the deposits, in order to make its mechanical characteristics better visible. The stabilizing pin has also been placed on its correct position, C) the digital "cleaning" of the Additional plate from its corrosion and the deposits. Also, the circular hole of the plate and the circular cross section of the gear shaft (probably the end of the shaft's edge), which is placed on the circular hole of the Additional plate, are presented. AMRP tomographies were processed by the authors. 1), 2), and 3) the corresponding bronze material reconstructions of the three layers of Fragment D, designed/constructed by the first author.*

In contact with the Circular plate, gear r1 is clearly visible (thickness about 1.5mm) and has been preserved in great condition. Its radius slightly varies as a result of the shrinkage deformation and the random cracks, Voulgaris et al., (2019b), between 16.7mm-17.2mm. This gear also has a square central hole. C. Karakalos, Price (1974); Wright (2005); Freeth et al., (2006); Freeth and Jones (2012), measured 63 gear teeth (61 teeth well preserved and two teeth missing). Around the perimeter of the gear stacked deposits of salts, mostly calcites, showing a strong absorption in X-rays, Voulgaris et al., (2018c) and petrified silt are located, following the shape pattern of the teeth. However, a large percentage of these formations have been peeled out of the teeth boundaries.

On this gear, the three pins which stabilized it on the Circular plate (Fig. 5) are sharply detected (stabilizing pins have also been detected on the gears c1/c2 and the l1/l2). Thus, gear-r1 and the Circular plate are fixed to each other and rotate with the same angular velocity as one body. It seems that this Circular plate is a base for the r1 gear, increasing its stability.

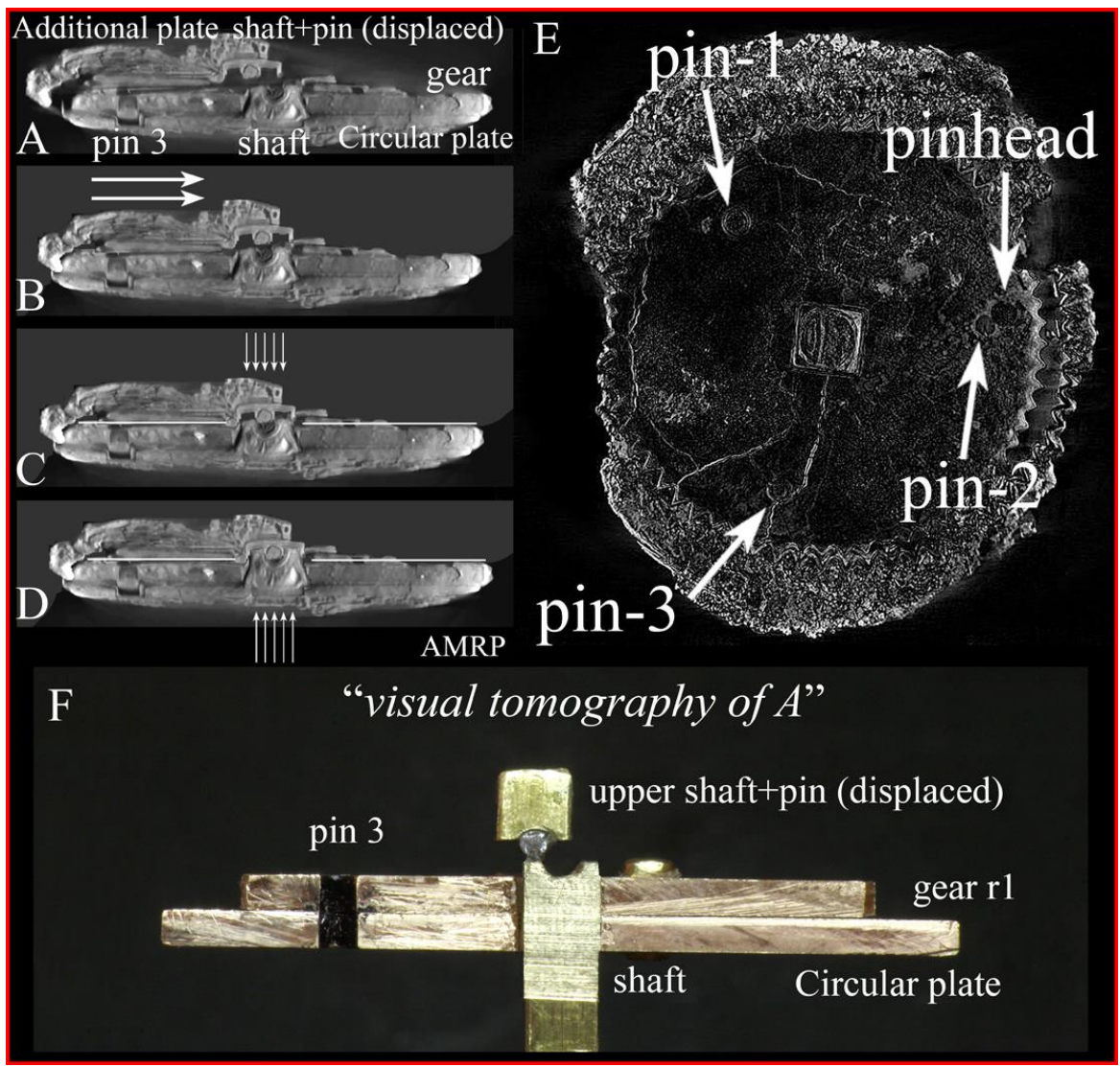

*Figure 6. The four stages for the relocation on the Additional plate's correct position and the broken r-shaft. A) The present position of the Additional plate and the shaft. B) Digital relocation of these two parts in order to align the broken shaft to the same axial direction with the central part of the shaft (i.e. the remaining shaft on the gear). C) Relocation of the shaft and the stabilizing pin in order for the pin to be level to the gear surface (white line). By the specific position of the hole on the main part of the shaft (i.e. below the gear surface), it is evident that the main part of the shaft (which is preserved inside the gear) is also displaced or shrunk due to deformation, Voulgaris et al., (2019b). D) A digital relocation of the main part of the shaft in perpendicular direction up to the stabilizing pin is presented. The relocation travel is about 0.4mm-0.5mm, (this procedure is also presented in Anastasiou Thesis (2014), without the fourth stage). Note that the perpendicular stabilizing pin is located between the gear-r1 and the Additional plate. The shaft's hole for the pin adaptation is located just on the upper surface of the gear (white lines). E) By subtracting the two AMRP CT's of Fragment D of different depth (CT-A surface of gear, CT-B, 1mm deeper) it is revealed that the head of pin-2 has been displaced. F) A "visual tomography of image A". The bronze Circular plate, the gear and the shaft were cut in half in order to simulate the X-ray CT of A. The upper part of the shaft was displaced as is the original. Bronze parts construction and images by the first author.*

Close to the gear surface and perpendicular to the gear shaft, a hole can be detected, in which the stabilizing pin of the gear was adapted. In this area, the shaft is broken. Above this hole, the AMRP tomographies reveal the stabilizing pin in contact to the broken shaft and a strangely shaped piece, the Additional plate. These pieces have been displaced from their original position, as is evident by the difference in position of the corresponding formations (gear shaft, hole and pin) Fig. 5,8. The Additional plate (thickness about 1mm-1.5mm) is partially preserved in a particular shape, Fig. 5,8. Visible on the Additional plate is a circular hole, in which the broken and displaced circular axis is adapted (in **Supplementary material** at the end, a further analysis regarding the position of Fragment D parts is presented and discussed).

The extended study of the AMRP Tomographies (original and also processed by the authors), did not give any proof for the existence of the three pins on the Additional plate, which could lead someone to consider that the Additional plate was fixed/stabilized on gear r1. The three pin's edges are clearly detected only between the Circular plate and the gear Fig. 6. Moreover, one of the three pins is totally out of the boundaries of the Additional plate Fig. 7.

Additionally, the existence of the perpendicular stabilizing pin of the shaft, between the gear and the Additional plate prevents any contact between the surfaces of gear r1 and the Additional plate Fig. 5B, and 9C,D.

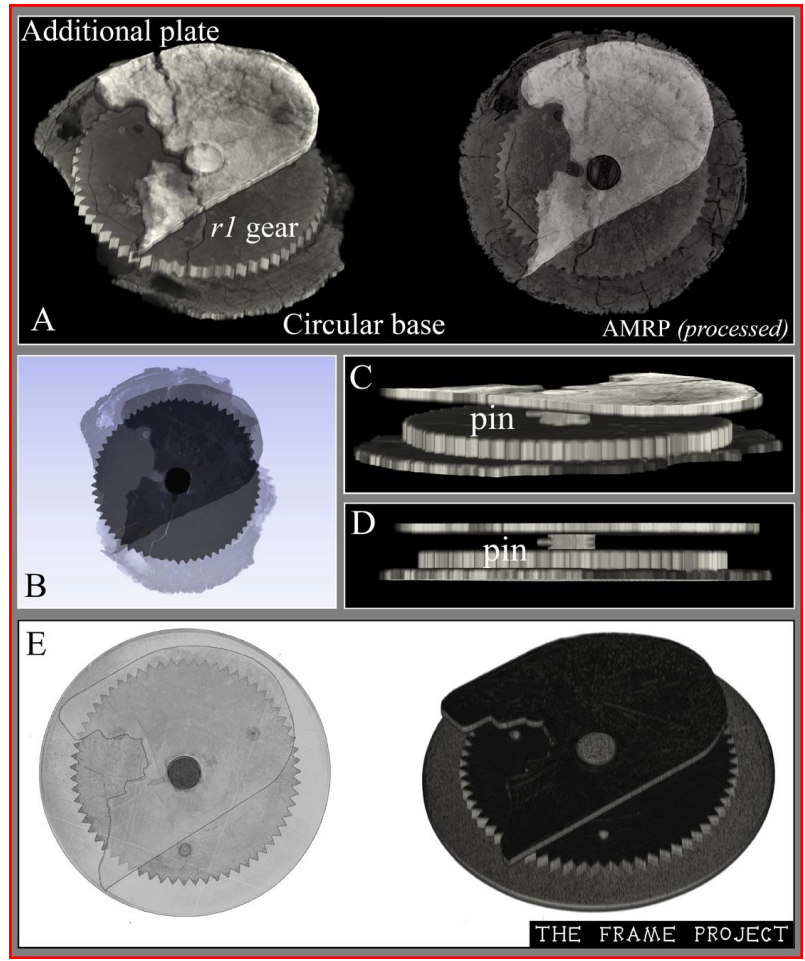

*Figure 7. A) A 3D representation using the 3D Slicer software, Fedorov et al., (2012), of the digitally reconstructed and "cleaned" AMRP tomographies of Fragment D, processed by the authors. B) The Fragment D is represented in transparent layers. C and D) Two side views in order to make the pin position visible. E) Left, the transparent and Right, the 3D representation of the "artificial radiographies" of the three assembled original bronze reconstructed parts, using the 3D Slicer software. Photos and processed images by the authors.*

The gap between r1-gear and the Additional plate is the thickness of the stabilizing pin, at least 0.9mm-1.0mm (in the bronze original a bit larger, taking into account the material corrosion and shrinkage). Therefore, the Additional plate is not in contact with the r1-gear and seems to be utterly independent from it.

Inside the circular hole of the Additional plate, the broken part of the gear shaft is also preserved, which at this point has a circular cross section Fig. 5,9. The circular cross section design means that during the rotation of the r1 gear and its shaft/axis the Additional plate does not rotate and is independent of the gear/shaft rotation.

The authors cannot find any realistic mechanical operation for this Additional Plate, other that it being a simple oblong plate - curved by one side (of which today only this part remains), acting as one out of two bearing points-plates for r-shaft (Fig. 14,16): most of the Mechanism axes/shafts, are supported on two opposite points-plates, Voulgaris et al., (2021). For example, the preserved shafts f, g, h, i, are supported between two parallel plates Middle and Back (this is also mandatory for the lost shafts m, n and o) and the d- shaft, between the Middle plate and the Ω-shaped Retention bar, Voulgaris et al., (2018c).

In Efstathiou et al., (2011); Anastasiou (2014); Basiakoulis et al., (2017), the Additional Plate is represented as fixed on gear-r1 and rotating with the same angular velocity. It is also suggested that this plate acts as a cam leading to the Equation of Time calculation. This consideration does not agree with the AMRP tomographies and the observations presented in this paper (the Additional plate is not in contact to the gear, it has not fixing pins, and it has circular hole - not square, i.e. it does not rotate with the gear, see also **Supplementary material**).

It is evident that if the ancient Manufacturer wanted to make the Additional plate rotate with the same angular velocity as the r1 gear, he should have at least made the hole of the Additional plate square, as well as the corresponding cross section of the r-shaft. Moreover, the stabilizing pin between the r1 gear and the Additional plate prevents the perfect immobilization of the Additional plate on the gear, even if it had a square cross section.

The existence and the specific position of the perpendicular stabilizing pin leads to the conclusion that the pin was adapted in order to stabilize the gear on the specific position, as such perpendicular stabilizing pins are preserved in shafts f, g, h and i. Therefore, the Circular plate is the base of gear r1 and from now on we call it Circular base. According to our design suggestion, the Circular base is in contact with the *Internal wooden casement* of the Mechanism, Voulgaris et al., (2019b). In this way, the diameter of the base is bigger than the gear's diameter and the stability of the gear is maximum.

The engraved letters "ME" are detected on three different places on Fragment D. Engraved letters have also been detected on two other gears of the Mechanism, on m1 (letter H) and b1 (letter N). These letters could be the assembly numbering of the parts (?? but not all gears have engraved numbers on them), or maybe a more imaginative scenario may be in play, e.g. the Manufacturer could have written his name or a phrase spread out on several parts of the Mechanism (?).

### 3.6. *The engaged gears b1-a1, the output on the a shaft*

The b1 gear is the Mechanism's larger gear, Freeth et al., (2006). In contrast to the e3 gear (the second larger), which is made by one circular plate, the b1 gear consists of several assembled parts, which form four radial bars and a ring with teeth and probably was constructed by the remained bronze parts-scrap. The rest surface area of the gear does not have any bronze material. The absence of the material resulted in the irregular and relatively strong, three-dimensional shrinkage deformation of the b1 gear, mostly by the long-time stand still of the Mechanism on the sea bottom, the effect of gravity and its abrupt dehydration after its retraction from the sea, Voulgaris et al., (2019b).

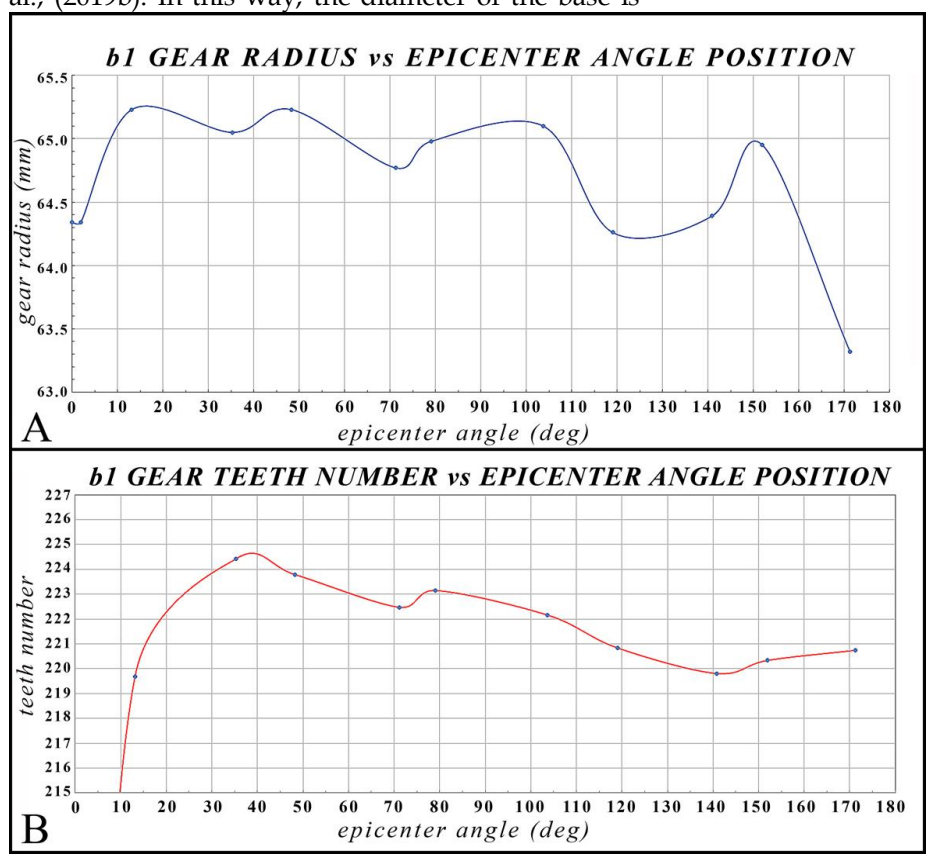

*Figure 8. A) The graph of the b1 gear radius vs angle position. Because the geometrical/mechanical centers of the two axes bin and bout, differ by about 0.4mm (as a result of the contractional deformation), the geometrical center of b1 gear central circular hole was selected as the center for the radius and epicenter angle measurements. B) The graph of the measured gear teeth number of b1 gear vs angle position. This graph more or less has similarities in the monotonicity to the above graph as a result of the radius variation (shorter radius results to smaller number of*

*teeth). Measurements by the Authors.*

Teeth number measurements of gear b1 (in its present condition) were made by C. Karakalos (223-226 teeth), D.S. Price (225 teeth), M. Wright (216-231 teeth) and Freeth et al., (2006) (223-224 teeth). An additional problem for the precise measurement of the teeth number is the difficulty for the detection of the actual teeth boundaries and tip position: many of the preserved gear teeth are filled with deposits (salts, calcites etc., as also observed on the r1 gear) that are following the shape of the gear perimeter, but at the same time they are altering or covering or erasing the triangular shape of the teeth. The teeth are worn out or destroyed or utterly missing at some areas. Moreover, the b1 gear radius varies by about ±1.1mm see the graph of Fig. 8, also in Freeth et al., (2006).

Firstly, the calculation of the total teeth number is achieved by measuring the preserved teeth and the corresponding epicenter angle (on their present condition and position). Secondly, by applying the equation: total gear teeth = (teeth number/corresponding epicenter angle)×360°. It is obvious that the corresponding epicenter angle is not the original-true angle, because of the gear's 3D shrinkage deformation, i.e. shorter volume and shorter dimensions than the original, Voulgaris et al., (2019b). The epicenter angle measured between the teeth tips on a part with shorter radius (i.e. shorter perimeter) results in a smaller calculated final teeth number (see Fig. 8). Apparently, the calculation of the mean value of the authors' measurements was avoided, because the statistic average values of the measurements do not approach the initial/original teeth position, as most (or all) of the teeth positions are displaced by the deformation. All these facts make the precise detection of the actual teeth boundaries and tip position, a difficult task.

Naturally, before the Mechanism's shrinkage, the original radius of the gear was larger. The authors' measurement of the gear teeth number is in the range of 219-225 teeth, concerning the present condition of the (deformed and shrunken) gear Fig. 8. The above leads to the conclusion that the original teeth number of the (bronze un-corroded/non-deformed/non-shrunken) gear must have been closer to the upper limit of this range.

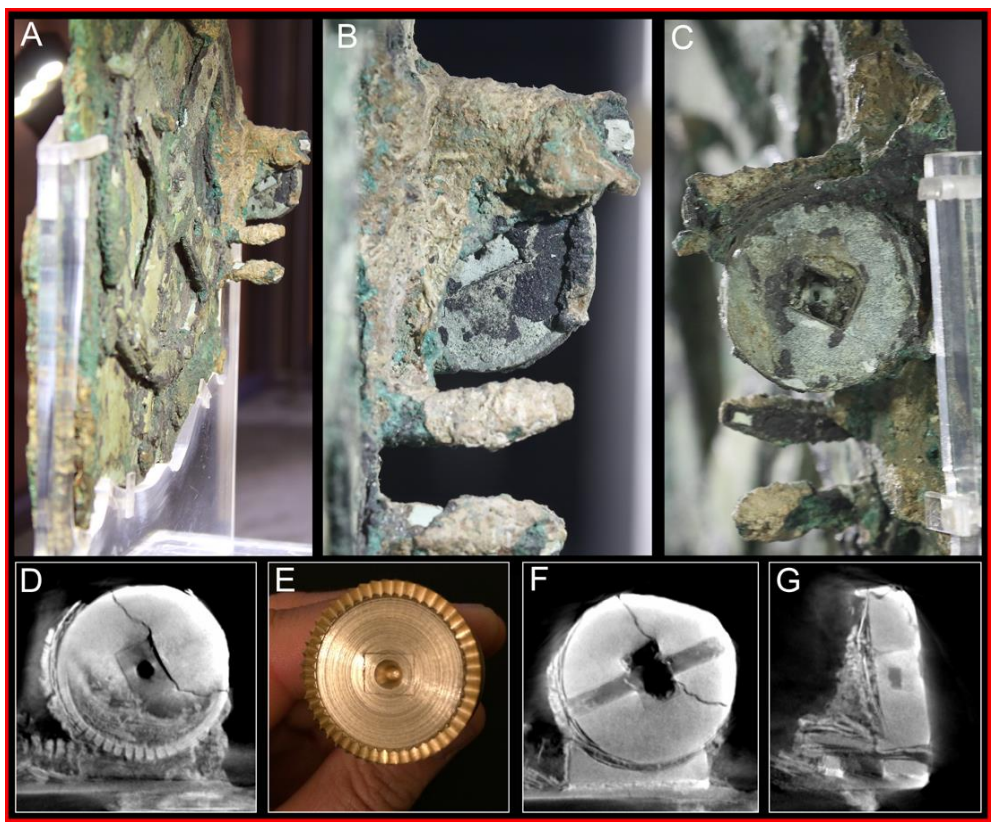

*Figure 9. A) The position of the contrate gear a1 relative to gear b1. B) Front face close-up of the contrate gear a1. The (conical) central hole on the shaft is slightly visible. C) Back face close-up of the contrate gear a1 (Credits: National Archaeological Museum, Athens, A. Voulgaris-Copyright Hellenic Ministry of Culture & Sports/Archaeological Receipts Fund Photos). D) AMRP right side view tomography of the gear a1. Some of the gear teeth are visible, and also the gear shaft with rectangular cross section and the (conical) central hole of the shaft. E) Bronze reconstruction of the a1 contrate gear, by the first author. F) The large pin, which immobilizes the a1 gear to its shaft, and the driver/nest of the*

*gear, are clearly visible. G) AMRP top side view tomography of the gear a1. The rectangular cross section of the large and thick immobilizing pin is visible.*

Gear a1 with (definite) 48 teeth, Freeth et al., (2006) is located on the right side of the Mechanism and is engaged with b1 gear Fig. 9. Gear a1 and its shaft rotate in a perpendicular direction to the rest gearing trains. The gear has a central rectangular hole, in which the poorly preserved rectangular shaft is adapted. On the shaft, a central conical hole is detected in AMRP tomographies, probably made by the ancient Manufacturer in order to process the material of the bronze shaft using his lathe, Voulgaris et al., (2018b) and (2019a). In AMRP tomographies, a relatively thick pin with oblong cross section, perpendicular to gear a1, is detected, Fig. 9F,G. This pin differs from the usual stabilizing pins, and it seems as an immobilizing pin, to fix a cylindrical bronze material (gear) to its shaft, probably before the material processing.

Between the a1 gear and the Middle plate, a part is detected which is stabilized on the Middle plate by using an adhesive material, an alloy of tin and lead, Voulgaris et al., (2018c). This part acts as a driver (nest) for the a1-gear in order to avoid the (probable) precession or displacement during its rotation, causing its disengagement with the b1 gear Fig. 9F.

Gear b1 is stabilized on gear b2 by four pins and rotates with the same angular velocity, Freeth et al., (2006), Voulgaris et al., (2019b). Gear b2 belongs to the main gearing of the Mechanism and its rotation represents the tropical year on the Mechanism. The larger b1 gear does not correlate to the main gearing of the Mechanism. Regardless of its teeth number or even if it did not exist, it would not affect the main gearing sequence of the Mechanism. Therefore, this gear was adapted by the ancient Manufacturer in order to introduce a new additional gearing, which started its movement by the engagement of the a1 contrate gear with b1 gear. The contrate gear transmits the movement in a perpendicular direction to the rest gearing axes direction. The presence of the a1 contrate gear leads to the conclusion that the ancient Manufacturer wanted to extend the gearing in a perpendicular direction, to be continued to the right side of the Mechanism. If his purpose was to continue new gearing's rotation in the same direction, he could simply use a 48 teeth gear instead of a contrate one.

In contrast to most of the other gearing trains, engaged in reducing/dividing ratios, the a1 gear/shaft rotation originates from a multiplying ratio: b1/a1 ≈ 4.6 rotations of a shaft per one rotation of b1(b2) gear-tropical year.

A number of researchers consider the contrate gear/shaft a1 as the "Input" of the Mechanism. For several significant mechanical and operational reasons the Input with a1 gear, presents mechanical and handling problems, Voulgaris et al. (2018b), see next Section.

## 4. THE LUNAR DISC/GEAR B3: THE IDEAL INPUT/DRIVING OF THE MECHANISM

### 4.1. *Lunar Disc Input vs crank-a1*

The assumption that the Input/driving of the Antikythera Mechanism comes from a1-gear by adapting a crank, was introduced by D.S. Price 1974. However, this common-sense assumption/though widely accepted does not necessarily have to be right.

Roumeliotis (2018) mentions that K. Efstathiou and M. Vicentini models (as also ours) cannot be operated at all or smoothly with the a1-crank. K. Efstathiou suggested that the handling of the Mechanism should be easier with the Lunar Disc.

Based in the construction of a functional model, in Voulgaris et al., (2018b) and below, we present three arguments proving that the input from a1-crank is highly doubtful, the rotation of the gears is not smooth and seamless, makes the handling of the Mechanism difficult. In Voulgaris et al., (2018b) and here we also answer why the input of the Mechanism should be from the Lunar Disc.

1) Suppose that the Mechanism's Input is from "crank a1-gear". Then, one complete rotation of the a1 crown gear will result in 2.85 rotations of the Lunar Disc (2.85 turns ~1026°), {(48/225) * (64/38) * (48/24) * (127/32) * (50/50) * (50/50) * (32/32)} ~ 2.85. Thus, if we rotate the "crank a1" by one tooth, the Lunar Disc will change position by 1026°/48 = 21.37° (about 71% of a zodiac dodecatemorion, about 21 subdivisions). This high speed rotation makes it difficult to aim the Lunar Disc pointer precisely at the Golden Sphere-Sun: E.g. if the Lunar pointer is 4° away from the Golden Sphere, then the "crank a1" needs to rotate by (4/21.37)= 0.1871 tooth fraction of a1 (i.e. rotating the crank by 1.4°), in order to bring the Lunar pointer directly to the Golden Sphere (New Moon, a critical position). This precision is too difficult to achieve if one takes into account the mechanical errors, backlash, friction, hand force and inertia, see Fig. 10. Moreover, this very fast rotation of the Lunar Disc prevents any successful aiming of the Lunar pointer to a selected Zodiac subdivision (e.g. to the 1st subdivision of a Zodiac constellation).

2) Roumeliotis (2018), who presents torque calculations on the shafts, writes: "…*The next candidate is gear b1, the one already assumed to be driven by a crown. Although this is a relatively good candidate, it gives a minimum torque about 5 times smaller than the one provided by gear e6 and about one third of the torque provided by gear d2. Thus, if the friction at each shaft or axle is larger than the estimated 0.2 N mm, driving the mechanism by gear b1 may be difficult.*"

3) In Voulgaris et al., (2018b), measured the kinetic energies considering the Input of the Mechanism by the crank-a1 ($E_{a1}$) and by the Lunar Disc ($E_{ld}$). The results revealed that $E_{a1}$ = 8.28 $E_{ld}$, i.e. 8.2 times more energy is needed to move the Mechanism gears with the "crank-a1" than to move it with the Lunar Disc.

The question is "*why the ancient Manufacturer designed and constructed an Input, which is less precise, difficult/challenging to handle, non-seamless and presents mechanical problems?*".

Moreover, if the Manufacturer wanted a better value of torque/better resolution in movement, he could easily decrease the teeth number of a1-gear, improving the precision of the Lunar Disc's aiming, but the problem of the friction and the inertia would remain for the following gears.

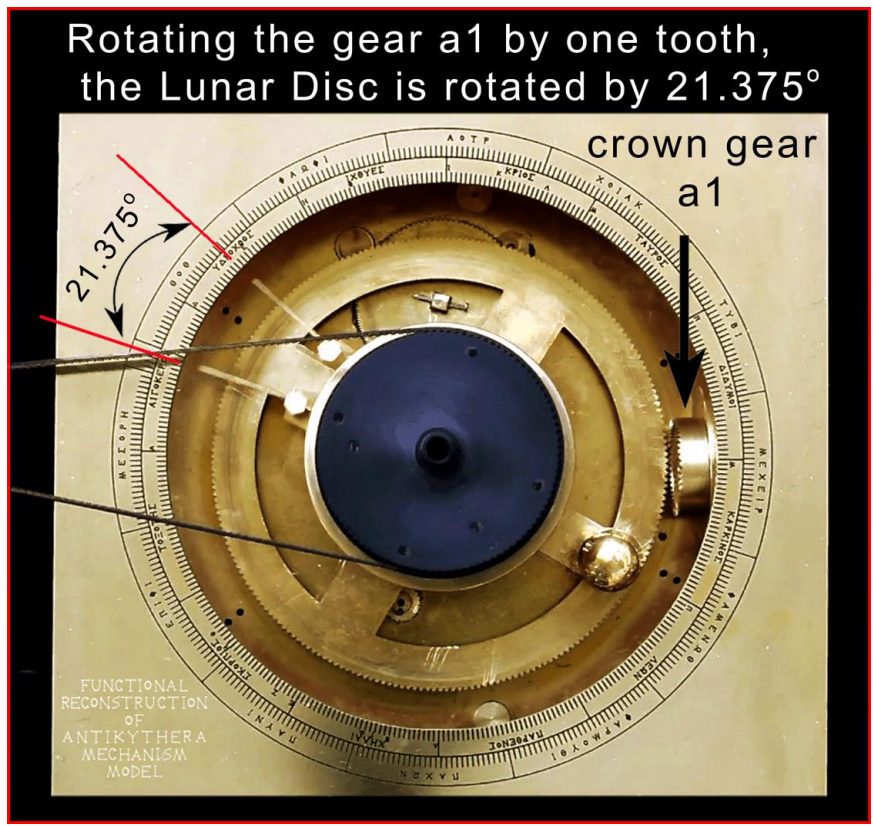

*Figure 10. Digital combination of two different images depicting the position difference of the Lunar Disc pointer, if gear-a1 (crank), rotated by one tooth. In this way, the aiming of the Lunar pointer to any desired point, is too difficult or even impossible.*

On the contrary, driving with the Lunar Disc offers easy handling, sufficient torque, and precise aiming to any point, even to a specific subdivision (1/365 of circle) of the Zodiac ring, Voulgaris et al., (2018a). In addition, it is fully functional and directly related to the units of the Saros and Metonic spirals, which are based on the Synodic cycle, Voulgaris et al., (2018b).

For all the reasons mentioned above, we do believe that the proper Input/driving of the Antikythera Mechanism is the Lunar Disc.

### 4.2. The engaged gears b1-a1, the output on the a shaft

Based on the arguments presented in the previous Section, the most proper, functional and easily handle Input/driving of the Antikythera Mechanism is the Lunar Disc.

1) This frees up the crown gear-a1 and its axis, which has an unknown (output) operation.

2) At the same time, there is an unplaced gear, the Fragment D, which also has an unknown operation.

3) During the Mechanism's Era four well-known Lunar cycles have been recorded, observed and used:

Synodic, Sidereal, Anomalistic and Draconic (223 Synodic cycles = 239 Anomalistic = 242 Draconic= 1 Saros cycle).

4) Three - out of four - lunar cycles are inherently present in the AM: Sidereal, Synodic, and Anomalistic (see *Section 3.1*), but not the fourth – very important – lunar motion, the Draconic cycle. As the Inscriptions of the Mechanism are partially and at some areas poorly preserved, and a large part is totally missing, it can be assumed that there might have been inscriptional evidence for a lunar node indicator on the Mechanism, which unluckily has not been preserved.

Therefore, it makes sense to investigate if the axis of gear-a1 and the unplaced gear of Fragment D could engaged in a Draconic gearing, which can cooperate with the rest, existing and actual gearing of the Mechanism.

## 5. RESULTS

### 5.1. Adapting the Fragment D on the Antikythera Mechanism

As explained in Section 3.4 there is no High Speed (High Resolution) gearing for the eclipse prediction based on geometrical (mechanical) calculations. At the same time, an unknown output of the Mechanism and an unplaced gear exists. Here follows a presentation of the rationale, the design, the gearing mathematical calculations, the use and the results of a new gearing, suggested by the authors, in order to achieve precise eclipse predictions based on geometrical calculations.

A new gearing can be introduced, to represent the second necessary parameter for the eclipse prediction: a gearing representing the Draconic cycle - the fourth lunar cycle - is needed (see *Section 3*). The output of the Draconic gearing is a Draconic pointer. This gearing train must be of High Speed (High Resolution), to extract results based on geometrical calculations.

Therefore, the Draconic gearing must be adapted at a position where the gears' speed is high. Moreover, a candidate position must allow the adaptation of the additional mechanical parts.

A proper position for this new-Draconic gearing can be found on the a-shaft-output, which starts with the engaged gears b1-a1. On the other edge of the a-shaft, the r1 gear is adapted Fig. 11. The mechanical design and the dimensions of the a-shaft allow the adaptation of the r-shaft/r1-gear. A hypothetical gear s1, engaged to r1-gear, is needed for the output of the Draconic gearing train. The Draconic pointer is adapted on the s-shaft Fig. 11,15.

One complete rotation of the Draconic pointer corresponded to one Draconic month. The measuring scale of the Draconic pointer depicts the Ascending and Descending Node-points. Two simple pins in up/down position, anti-diametrically placed and stabilized on the External Wooden decorative casement, Voulgaris et al., (2019b), represent the two Nodes: one gold pin for the Ascending Node and one silver pin for the Descending Node. On either side of each Node-pin, there must also be an arc-shaped bronze strip, depicting the Zone of Eclipses/Ecliptic limits, about ±17° for each Node Fig. 14,17 (these ecliptic limits are approximate values and will be recalculated as the research progresses). A Draconic month is completed whenever the Draconic pointer returns to the same Node-pin, after one rotation.

For the gear teeth calculations, the lunar cycles of Saros period (Table I) were selected as the most proper period for the eclipse prediction.

One Saros period of 223 Synodic months = 242 Draconic months. Therefore, on the Antikythera Mechanism, 242 complete rotations of the Draconic pointer are equal to 223 Synodic rotations of the Lunar Disc (i). On the Metonic gearing, 235 Synodic rotations = 254 Sidereal rotations or 1 Synodic rotation of the Lunar Disc Input = 254/235 Sidereal rotations (ii).

Therefore, 242 Draconic rotations = 223 * 254/235 Sidereal rotations (iii). Applying this equation on the Antikythera Mechanism gearing:

$\{223 * (254/235)\} * (b3/e1) * (e6/k2) * (k1/e5) * (e2/d2) * (d1/c2) * (c1/b2) * \{(b1/a1) * (r1/s1)\}$ = 242 rotations of the Draconic pointer (iv), therefore

$(b1/a1) * (r1/s1) = 13.42223271$ (v).

For a1 = 48 teeth (definite) and r1 = 63 teeth (definite), the equation (v) becomes

$b1/s1 = 10.22646302$ (vi).

For a gear teeth number of b1 = 225 (see *Section 3.6*), the equation (vi) yields

$s1 = 22.00174$ teeth, rounded to 22 teeth for the gear s1.

From the above calculations it follows that for one rotation of the b1 gear (one Callippic tropical year of 365.25d), the Draconic pointer rotates 13.42329545 times (Draconic months), so 365.25d/13.42329545 = 27.2101587d/Draconic month, instead of 27.21218683d (resulting from the values of the Saros cycle presented on Table I). The gearing exhibits a phase difference-error of -0.00202809d/1 Draconic pointer rotation, i.e. about -2.9 min/Draconic month. Without a doubt, such an error is too small compared to the mechanical errors and manufacturing imperfections (e.g. mismatches between the gear teeth shape and eccentricities) of the Mechanism gearing, Edmunds (2011).

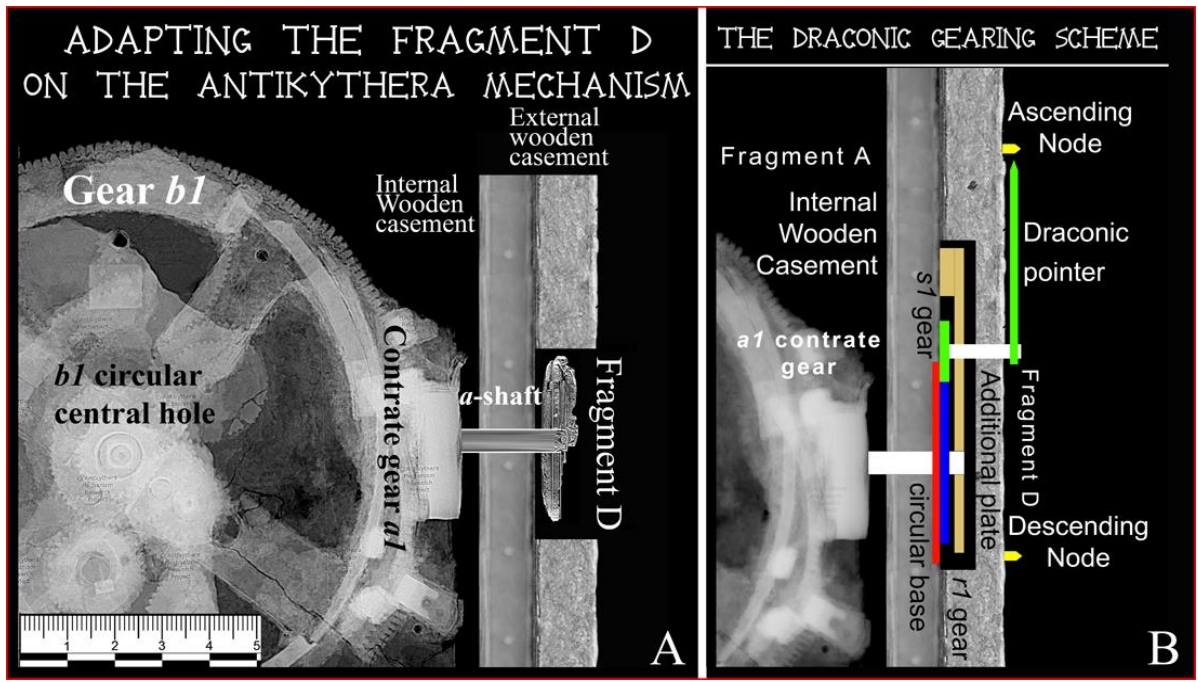

*Figure 11. A) Same scale, digital adaptation of Fragment D on Fragment A. The a-shaft (poorly preserved today), is connected to the r-shaft of Fragment-D. B) The Draconic gearing scheme-Fragment D, is adapted on the Internal Wooden Casement (Voulgaris et al., 2019b), on the right side of the Antikythera Mechanism. The a-shaft, the Circular base, the (hypothetical) gear s1, the Draconic pointer and the two Node-pins are presented. The Circular base enhances the stabilization of the a1 and the r1 gears. AMRP radiography (Fragment A) and tomography (Fragment D) were processed by the authors.*

## 5.2. A further analysis and discussion for Fragment D parts

By studying the way of the gears' stabilization of the Mechanism we detected some crucial technical characteristics, introduced by the ancient Manufacturer. One face of **all of the preserved gears** is in contact to another fixed mechanical part i.e. the gears are based somewhere in such a way that stability is achieved: As the gears of the Mechanism are relatively thin, around to 2-3mm, when rotating they shake a lot, because the gear cannot continuously stay perpendicular to its axis (the stabilizing perpendicular pin does not offer a perfect stabilization), Fig. 12A, resulting in the gears' disengagement, see Voulgaris et al., 2018b. This error is generally named *circular* or *axial runout* (https://www.youtube.com/watch?v=mqqda5FFB9o&t=4s). The authors called this effect "the *Libration effect* of the Antikythera Mechanism gears", Fig. 12A, as this motion seems like the lunar Libration (by observing the Lunar Equator). In order to eliminate this critical mechanical problem of the *gears' Libration*, the ancient Manufacturer either places the gears directly in contact to the Middle plate, face to face (gears c2, d1, l2, m1, e5, k1, e1, b3), or he adapts supportive spacers in "C" shape (gear b1, d2, e3), or stabilizes the gears on cylindrical, square or otherwise shaped spacers, having an adequate contact surface (gears b2, f1, g1, h1, i1, m3). The gears' rotation is impossible without any supporting system. Therefore, each gear needs its supporting base.

Freeth et al., (2021) suggests that Fragment D is the output for the hypothetical gearing of planet Venus. In their suggested design they present the Additional plate as the base of gear r1 (i.e. in inverse/top-bottom position, relative to the design in this work). In Figure 4p of Freeth et al., (2021) and in Figure S14 of their *Supplementary Information (also in Supplementary Information 1*, of their video at 0:28 – 0:34), the stabilizing pin is presented at a different position and not between gear-r1 and the Additional plate. According to AMRP tomographies, the pin is displaced and the hole for the pin adaptation is clearly located right after gear-r1, see Fig. 13 and not above the Additional plate, as is presented by Freeth et al., (2021). In authors' **Supplementary material**, an analysis of the parts position according to Freeth et al., (2021) suggested design of Fragment D is presented and discussed.

Let us analyse the functionality of Fragment D by rotating it 180°, i.e. the Additional plate as the base of the fragment, on top of it the stabilizing pin, then gear-r and finally the Circular plate on top, Fig. 12 B,C,D.

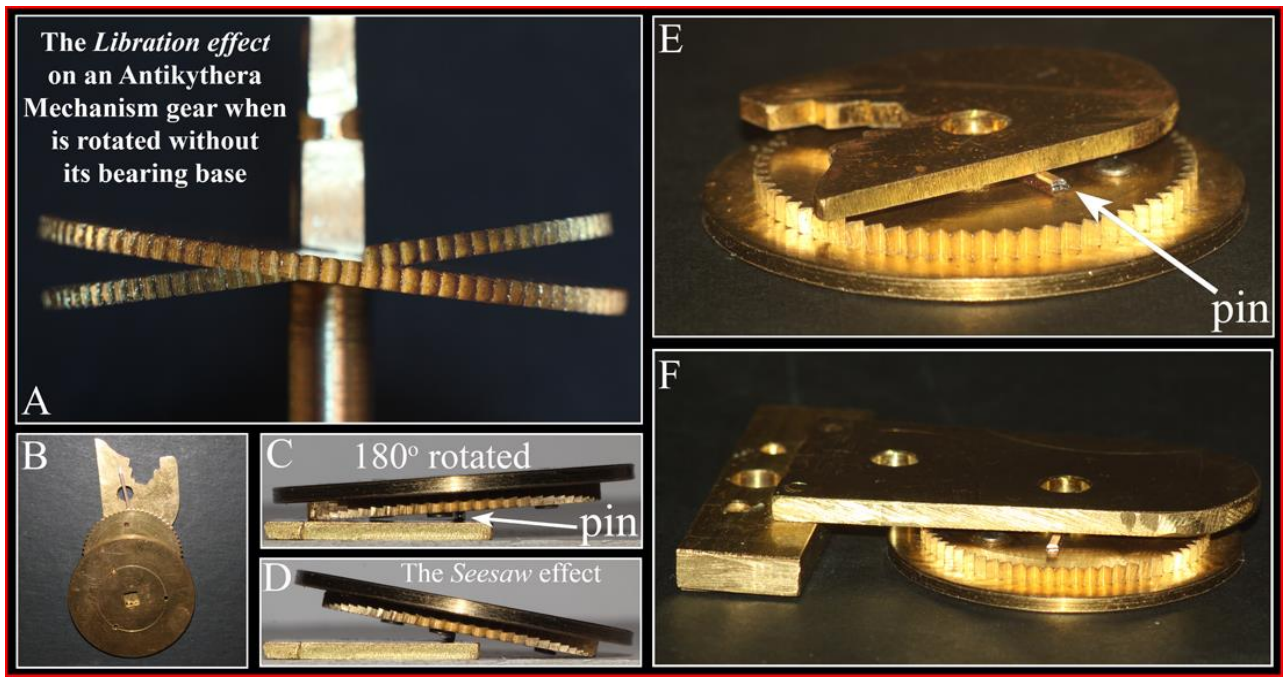

*Figure 12. A) The "Libration effect" on an Antikythera Mechanism gear, which is rotated without its base. B) Considering that the Additional plate is the base of gear r1, (inverted position of Fragment D). C,D) In the inverted position of Fragment D, the "Seesaw effect" appears in gear r1 during its rotation. Because of the pin's existence, the gear cannot remain continuously perpendicular to its shaft, leading to its disengagement with any other gear. As gear r1 is not based anywhere it cannot be functional. E,F) On the contrary, the suggested design in the present work, offers high stability and functionality of the gear and it is mechanically acceptable. This design follows the design of the today preserved gears adaptation pattern and the constructional style of the ancient Manufacturer. Bronze parts and images by the authors.*

In this design, contact of gear r1 with a base (plate or spacer) is totally missing: Between the hypothetical base of the system (the Additional plate) and gear-r1, there is the pin (see Fig. 12 B, C, D and Fig. 13), which prevents any contact of the gear to its hypothetical base. The pin destroys any stability of the gear and the gear presents a strong shaking during its rotation, like a Seesaw motion, because it is not supported by any part, (see Fig. 12 C, D). Therefore, the rotation of the fragment by 180° raises the question "*why did the ancient Manufacturer adapt the pin between the gear and its hypothetical base, creating significant stability problems on this system?*".

Moreover, the thinner part is considered as the base of the system and the moving parts are thicker.

Based on our experience coming from the Mechanism bronze parts construction, assembly, handling, on the observations of the present work and according to our opinion, the inverse position of Fragment D is mechanically unorthodox, does not concur with the constructional characteristics inferred by the preserved gears, and its functionality is quite doubtful.

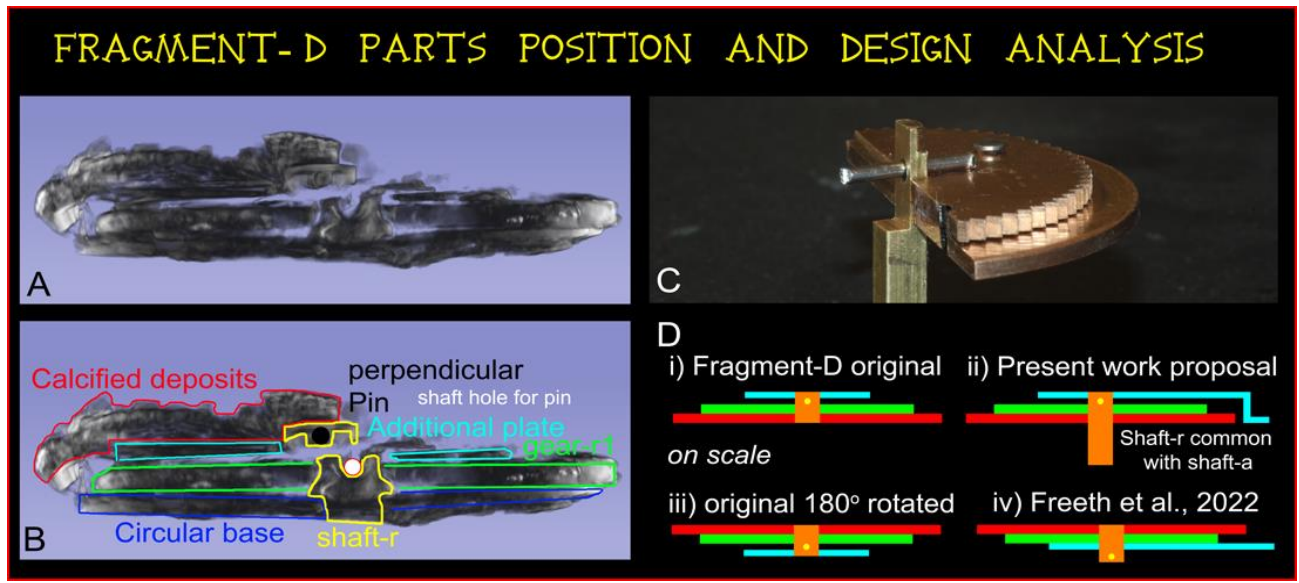

*Figure 13. A) AMRP 3D reconstruction of Fragment D was cropped at the area of the gear-r1 center (via 3D Slicer software). B) The positions of the Fragment D parts are marked in different coloured lines. The broken upper part of shaft r (yellow line) and the stabilizing pin (black dot) have been displaced. The rest parts of the fragment, the Circular base (in blue), gear-r1 (in green) and shaft-r, have remained on their original position. C) A restored "visual tomography of image A": the perpendicular pin and the broken part of shaft-r are placed in their original position. The pin is in contact to the upper surface of the gear. Bronze parts construction and images by the first author. D) i) The original design of Fragment D, after the repositioning of the displaced parts. ii) Fragment D proposed design in the present work. iii) Rotating scheme i by 180°. iv) Suggested design by Freeth et al., (2021) for the hypothetical gearing of Venus output. The position of the stabilizing pin is presented as the last/outer part of the Fragment D and does not agree with the AMRP tomographies (see Supplementary material).*

## 5.3. The "scenic" operation of the Draconic gearing on the Antikythera Mechanism

Introducing the Draconic gearing on the Antikythera Mechanism, the operation of the high-resolution eclipse prediction based on the geometry is achieved. The specific position of the Draconic pointer on the right side of the Mechanism Fig. 14, also has a special scenic operation: as the user rotates the Lunar Disc when the lunar pointer aligns with the Golden Sphere-Sun, signifying the New Moon phase (or, in the opposite direction – the Full Moon phase), he can also easily look at the right side of the Mechanism to see if the Draconic pointer is located within the Ecliptic limits (arc bronze strip). If this is the case, he can be certain knows that a solar (or lunar) eclipse will occur. Afterwards, he can turn the Mechanism to the other side (Back plate) and by observing the cell that the Saros pointer aims at, he can read the eclipse information about the time of the eclipse event and the corresponding Metonic month in which the eclipse occurs. If the Draconic pointer aims anywhere out of the Ecliptic limits, the corresponding Saros cell is absolutely empty (see Fig. 14, 17).

Someone may wonder why the ancient Manufacturer did not place the Draconic gearing in a dial display close to the Saros dial of the Back plate. On the Back plate there are gears with quite slow rotation (Metonic pointer: 1 turn/3.8 years, Saros pointer: 1 turn/4.5075 years, Athletic Games pointer: 1turn/4 years). So it is difficult to connect a gear with fast rotation like the Draconic pointer which completes 13.422 turns/year (about equal to Sidereal 13.368 turns/year), with a very slow gearing train. Therefore, the ancient Manufacturer tried to find for the Draconic gearing a place with fast rotation gears.

Someone also may wonder why the ancient Manufacturer did not place the Draconic gearing close/around to the Lunar Disc gearing: The Lunar Disc completes 13.368 turns/year and the Draconic pointer 13.422 turns/year. For the ratio Draconic/Sidereal≈ 1.004039, an accepted approximation of two engaged gears is 249 and 248 teeth. These gears should have a diameter of around 116mm, which needs a large space (232mmX116mm). The Manufacturer could also make a combined gearing by several gears of (6X83)/(16X31) a complex combination (83 and 31 are prime numbers and a gear with 6 teeth is a difficult and doubtful construction). Moreover, additional space is needed for the Draconic pointer and scale.

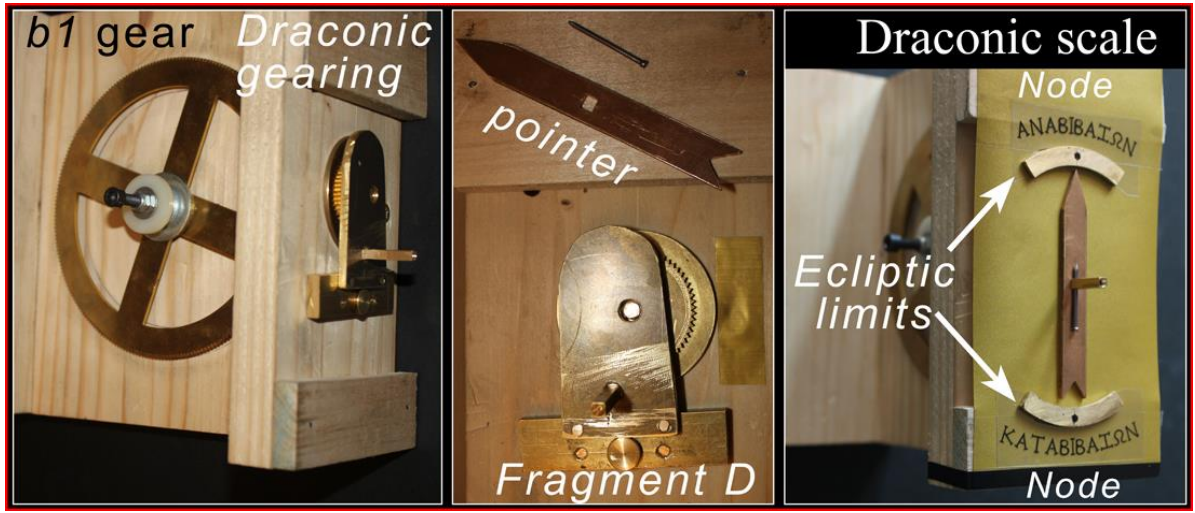

*Figure 14. Bronze reconstruction of the Draconic gearing. The b1 gear, the reconstruction of Fragment D/gear-r1, the Draconic pointer and the Draconic scale, with the two Nodes (Ascending and Descending) and the corresponding ecliptic limits, are visible. In this design, when the Draconic pointer aims to a Node, it is parallel to the Mechanism's Ecliptic, i.e. the Zodiac ring plane of the Front Dial plate.*

Regarding the letters "ME", appearing three times on the parts of Fragment D, one could assume that they are the headings of the words "ΜΗΝ ΕΚΛΕΙΠΤΙΚΟΣ" or "ΜΗΝ ΕΓΛΕΙΠΤΙΚΟΣ" (Ecliptic Month), which is related to the three Draconic gearing parts (as a rough/fast coding of the specific parts by the Manufacturer). The Ecliptic (ΕΚΛΕΙΠΤΙΚΗ) is referred to in the Mechanism Back cover inscriptions as "…ΟΙ ΕΓΛΕΙΠΤΙΚΟΙ ΧΡ[ΟΝΟΙ…]", times (?) of eclipses, Bitsakis and Jones (2016).

### 5.4. The variable velocity of the Moon - pin&slot gearing motion, visible on the Draconic pointer.

By starting the Mechanism with the Lunar Disc-Input, it is obvious that the variable lunar velocity produced by the *pin&slot* invention in gears k2/k1 is transmitted via the gearing to the last gear s1/pointer of the Draconic gearing. Therefore, the Draconic pointer has a variable angular velocity, and each time crosses a Node of the Draconic scale at different velocity. This also occurs in reality, as the four lunar cycles are directly affected by the motion variability of the Moon.

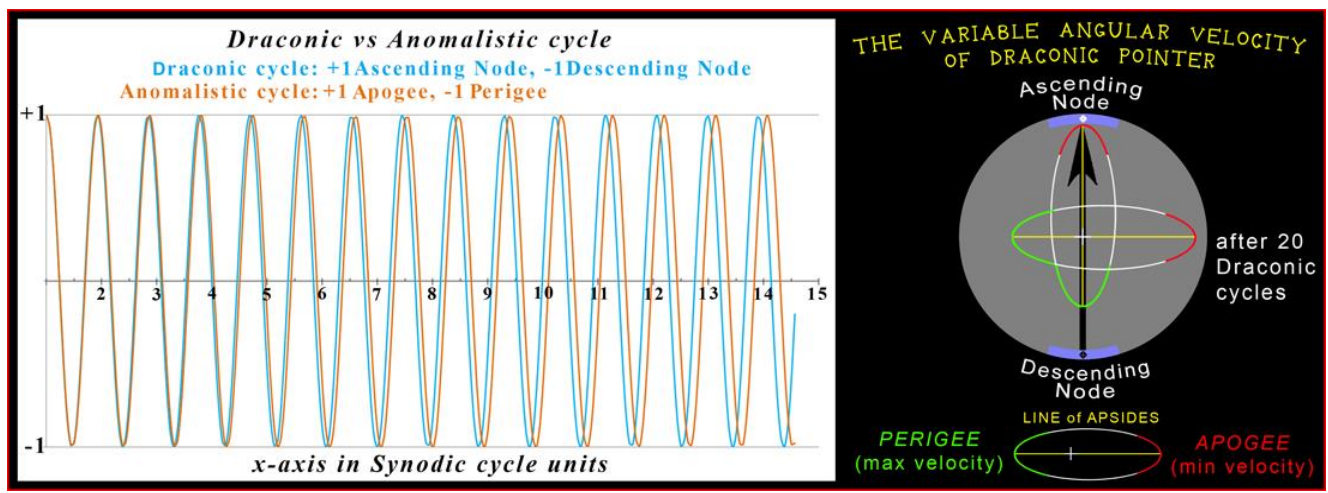

*Figure 15. Left, the harmonic graph of the Draconic and Anomalistic cycles, assuming that they start on the same phase (Draconic in Ascending Node, Anomalistic in Apogee) and during the New Moon phase. The units of the x-axis (time) are in Synodic cycles. Right, the representation of the variable angular velocity of the Moon via the pin&slot invention, is transferred to Draconic pointer (black arrow) rotation. The Draconic pointer crosses the Ascending Node with its minimum velocity (Moon at Apogee). After 20 Draconic cycles, the Draconic pointer crosses again the Ascending Node having the mean angular velocity of the Moon. After one Sar (half Saros) equals to 121 full rotations of Draconic pointer, the angular velocity of Draconic pointer is maximum (Moon at Perigee).*

The rotation period of the Draconic output-pointer is about equal to the Lunar Disc Sidereal rotation period: from equations (i) and (ii) it follows that 1 Saros = 242 Draconic cycles ≈ 241.029 Sidereal cycles i.e. 1 Sidereal rotation (360°) of the Lunar Disc ~ 1.004028 Draconic rotations (~361.45°), i.e. the Draconic pointer rotates a bit faster than the Lunar Disc pointer Fig. 15.

On the Saros cycle, 242 Draconic months are also equal to ≈239 Anomalistic cycles i.e. 1 Anomalistic month (from Apogee to the next Apogee) ~1.01255 Draconic months. This means that the Line of Apsides (the line connecting the points of min and max lunar velocity) delays each Draconic rotation, i.e. it rotates in the opposite direction relative to the Line of Nodes.

On the Antikythera Mechanism, this change is visible on the Draconic pointer. The Draconic pointer rotation presents a variable angular velocity, and the max-min velocities (Line of Apsides) change their position relative to the Line of Nodes. The imaginary Line of Apsides slowly rotates in the opposite direction (CW) relative to the Draconic (CCW). This implies that the Draconic pointer returns to the same Node-pin, each time by a different velocity, as it also happens every time the Moon approaches a Node, Fig. 15.

In Freeth et al., (2021), the suggested gearing design for the pointer of lunar Nodes is driven (via the gear b1) from crank-a1. In this design the pointer of Nodes is rotated in constant angular velocity. As the Draconic cycle is a (variable) motion of the Moon (produced by the Anomalistic cycle), it is hard to explain why the fourth lunar cycle is correlated with rotating gears having a constant angular velocity.

## 6. DISCUSSION

In this work, an ideal position and function for the enigmatic Fragment D is presented, taking into account the present condition and the preserved parts of the Antikythera Mechanism. According to the authors' opinion, the Draconic gearing existence offers the necessary precise eclipse prediction/calculation of the Mechanism. The authors tried to correlate, using the minimum number of necessary hypotheses, an unplaced gear with an unknown gear train output of the Mechanism, without adding any too hypothetical or theoretical parts and scenarios. For this attempt, the dimensions of the preserved parts were taken into account. The simple design/construction of the Draconic gearing includes the use of the three existing gears (b1, a1, r1) as well as a new gear (s1). No additional, hypothetical complex parts and engraved scales were needed.

This additional gearing train improves the instrument's efficiency, which can now perform precise eclipse predictions/calculations. The adaptation of the Draconic gearing/cycle on the Antikythera Mechanism presents a complete representation of the four lunar motions that were well-known and studied in the Hellenistic era. The authors strongly believe that the ancient Manufacturer of the Antikythera Mechanism took into account all four integrated lunar motions when he designed his creation.

Summarizing:
1) A realistic and relevant to the Antikythera Mechanism operation for the unknown a1 output was found,
2) The multiplying rotation of the a1-gear leads to a gear train output a few times faster in rotation than the tropical gear,
3) A mechanically accepted position and role for the unplaced Fragment D/gear-r1, was detected,
4) The existence of the two other parts of Fragment D was justified,
5) The mathematical calculations of the Draconic gearing are highly accurate,
6) There is adequate space for the Draconic train adaptation at the right side of the Mechanism, and this specific position assists the Mechanism user during the operation,
7) The teeth number of the b1 gear was specified,
8) The existence of the specific high-speed gear train offers the geometrical calculations needed for the eclipse events prediction/high resolution calculation of the Antikythera Mechanism, improving the accuracy of the eclipse predictions,
9) The addition of the Draconic gearing on the Mechanism introduces the two mandatory lunar cycles for the eclipse prediction, Synodic and Draconic,
10) The variable angular velocity produced by the *pin&slot* invention, is transformed to the Draconic pointer, as this occurs in all of the lunar cycles in reality,
11) Finally, by introducing the Draconic gearing on the Antikythera Mechanism, incorporating on the Antikythera Mechanism gearing all four, well-studied in antiquity, interrelated lunar motions, is achieved, Fig. 16.

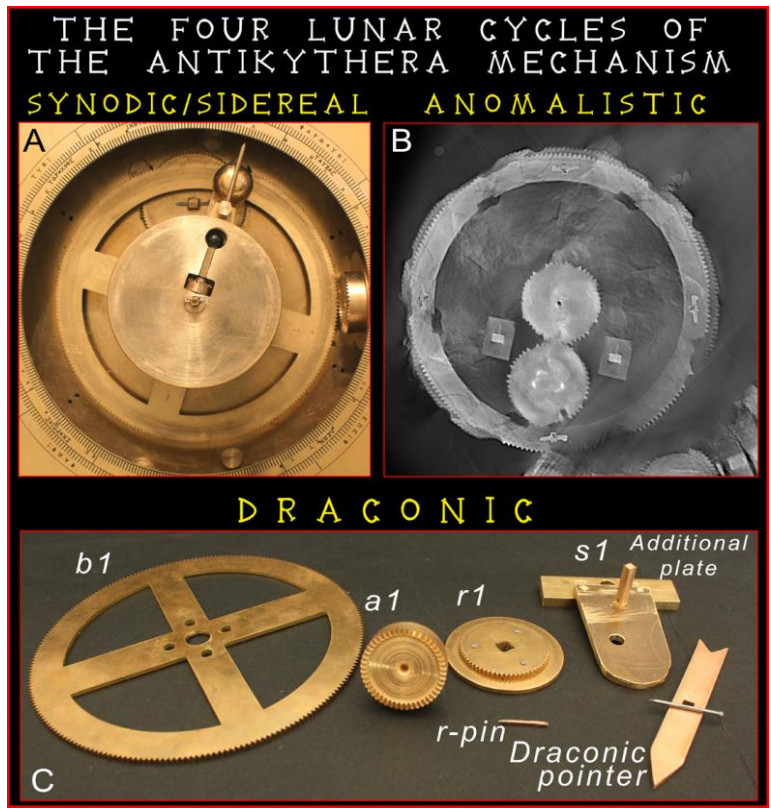

*Figure 16. A) A comprehensive view of the four lunar cycles represented on the Antikythera Mechanism: the Synodic/Sidereal cycle in a bronze functional model reconstructed by the authors. B) The Anomalistic cycle via pin&slot, in AMRP combined tomography and a digital completion of the lost half gears e3, e4, e6 and k2, processed by the authors. C) The (suggested and necessary according to the authors' opinion) Draconic gearing parts: b1, a1, r1, and s1 gears, the completed Additional plate and the Draconic pointer, constructed by the first author in bronze.*

## 7. CONCLUSION

Although someone should be sceptical when introducing hypotheses regarding the Antikythera Mechanism, the authors believe there is a large number of reasons justifying the existence of this new operation/gearing. According to the authors' opinion it is hard to explain why the ancient Manufacturer chose to include in his creation the Synodic, the Sidereal and the Anomalistic lunar cycles, but to exclude the Draconic cycle, which is critical for the eclipse prediction. Additionally, they cannot find any mechanical problem or non-relation or teleological reason, in order to doubt or reject this new gearing. The Draconic gearing is the key to predict the eclipse events (and their hours occurred) by only using the Antikythera Mechanism, without any other external information, not directly related to the Antikythera Mechanism.

The Fragment D(*raconic*) of the Antikythera Mechanism can be correlated to the four lunar cycle, the Draconic cycle. The impact of this correlation completes the Antikythera Mechanism representation of the four lunar cycles, without the assumption of a large number of components: just by adapting of the actually existing Fragment D and only one hypothetical small gear, the fourth Lunar cycle takes its place on the Mechanism.

According to our study, which is in progress, we have evidence to believe that the ancient Manufacturer constructed his creation in order to mechanically detect (predict) the unknown eclipse events and the times occurred by introducing the four lunar cycles on the Antikythera Mechanism and without any other external information that not related to the Mechanism, Fig. 17.

Including the fourth lunar cycle, the Draconic gearing and pointer to his construction, the ancient Manufacturer had the ability to predict/(self)calculate "*automatically*" the unknown eclipse events. Based on the self-calculated predictions and without any previous information that did not come from the Mechanism's calculation, he engraved the eclipse events on the (blank) cells of the Saros spiral, after the assembly of his creation.

The ancient Manufacturer "*trained*" his machine, in order to make astronomical event predictions via a "clear" mechanical procedure.

This distinctive operation of the Antikythera Mechanism may well have been the very first application of the *machine learning* concept in the history of mankind.

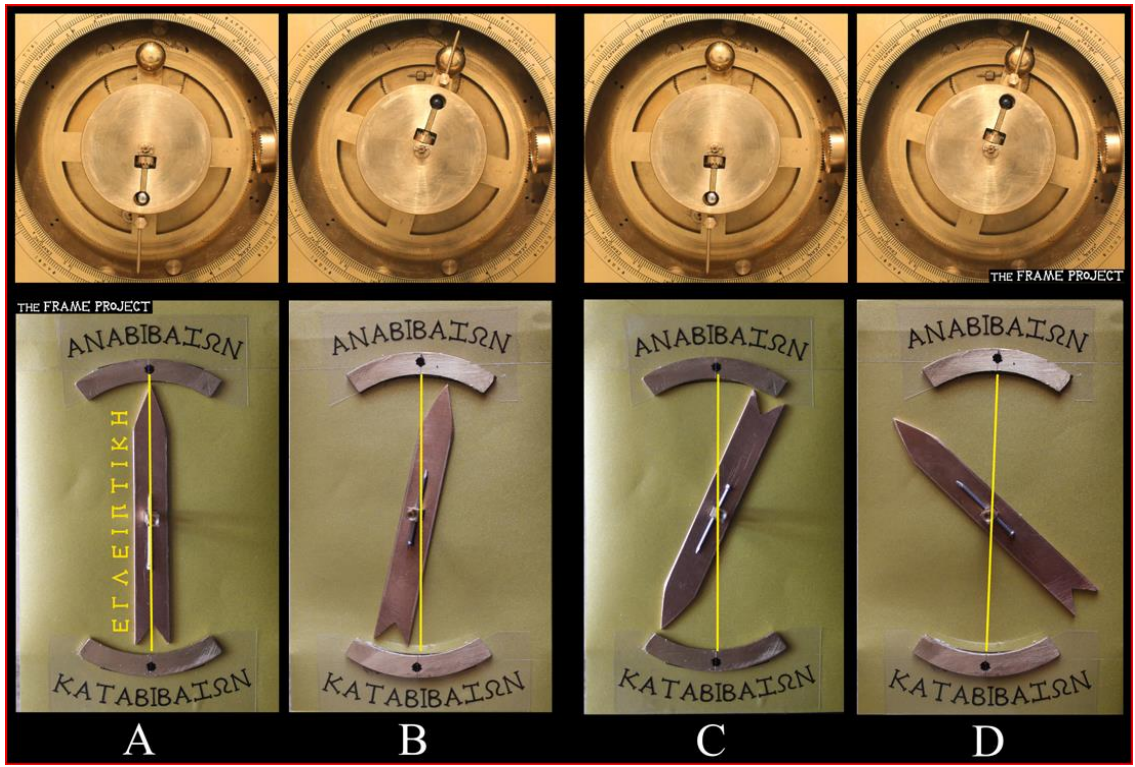

*Figure 17. Top, The front dial plate of the functional model by the authors. The Lunar Disc and the Golden sphere-Sun are visible. Bottom, The Draconic scale consists of the Draconic pointer, the Ascending (ΑΝΑΒΙΒΑΖΩΝ) and the Descending (ΚΑΤΑΒΙΒΑΖΩΝ) Nodes (black dots) and the two ecliptic limits (in arbitrary angle magnitude, in the present work), are depicted. The yellow line shows the direction (plane) of the Zodiac ring (Ecliptic-ΕΓΛΕΙΠΤΙΚΗ), located on the Mechanism's Front dial. The Draconic pointer direction depicts the lunar position relative to the Ecliptic plane (on, above or below the Ecliptic plane). The phase positional correlation between the Lunar Disc pointer relative to the Golden sphere-Sun and the Draconic pointer relative to the Nodes and the Ecliptic limits define an eclipse possibility event: A) The Full Moon is located on the Ascending Node, a total lunar eclipse occurs. B) The New Moon is located faraway from the Ascending Node, a total or annular solar eclipse occurs, visible faraway from the Earth's Equator. C) The Full Moon is located just on the ecliptic limit boundary, a marginal partial or penumbral lunar eclipse occurs. D) The New Moon is located out of the Ecliptic limits, no solar eclipse occurs.*

*Note that during the total solar eclipse of December 4th 2021 the New Moon was located faraway from the Descending Node, and it was much closer to the ecliptic limit. For this reason, this eclipse was only visible from locations at very high southern latitudes, Antarctica and nearby areas http://xjubier.free.fr/en/site_pages/solar_eclipses/TSE_2021_GoogleMapFull.html, http://nicmosis.as.arizona.edu:8000/ECLIPSE_WEB/TSE2021/TSE2021WEB/EFLIGHT2021.html, https://earthobservatory.nasa.gov/images/149174/antarctica-eclipsed*

## AUTHOR CONTRIBUTIONS

Writing - original draft preparation, A.Voulg. and C.M.; methodology A.Voulg.; software, C.M.; investigation, A.Voulg. and C.M.; Image process and data visualization A.Voulg. and A.Vossin; x-ray CTs process, A.Vossin. and A.Voulg. All authors have read and agreed to the published version of the manuscript.

## ACKNOWLEDGEMENTS

We are very grateful to Professor Xenophon Moussas (National and Kapodistrian University of Athens, Greece) providing us the full AMRP X-ray Volume Raw data of Antikythera Mechanism. John Hugh Seiradakis (Aristotle University of Thessaloniki, Greece) was providing us AMRP X-ray CT's of some of the large fragments, before he passed away on May 2020. Thanks are due to the National Archaeological Museum of Athens, Greece, for permission to photograph and study the Antikythera Mechanism fragments. We would also like to thank Prof. Thanasis Economou (Fermi Institute-University of Chicago, USA) for his suggestions. We also thank the two anonymous referees for their comments.

# REFERENCES

Anastasiou M., (2014) The Antikythera Mechanism: Astronomy and Technology in Ancient Greece. Aristotle University of Thessaloniki, *Phd thesis* (in greek with english summary).

Anastasiou, M., Seiradakis, J. H., Carman, C.C. and Efstathiou K. (2014) The Antikythera Mechanism: The Construction of the Metonic Pointer and the Back Dial Spirals. *Journal for the History of Astronomy*, 45, pp. 418–41.

Anastasiou, M., Bitsakis Y., Jones A., Steele J. M. and Zafeiropoulou M. (2016b) The Back Dial and Back Plate Inscriptions. In Special Issue: The Inscriptions of the Antikythera Mechanism. *Almagest* 7(1), pp. 138–215.

Antarctica Eclipsed, 2021. https://earthobservatory.nasa.gov/images/149174/antarctica-eclipsed

Antikythera Mechanism Research Project, http://www.antikythera-mechanism.gr

Barbieri, C. (2017) *Fundamentals of Astronomy*. Florida, CRC Press, Chapman & Hall.

Basiakoulis, A., Efstathiou, M., Efstathiou, K., Anastasiou, M. and Seiradakis, J. H. (2017) Ancient Technology and the Correction of the Time Shown by the Sun watches, *Proceedings of the 6th International Conference on Manufacturing Engineering "ICMEN"*. 6th International Conference on Manufacturing Engineering "ICMEN". Thessaloniki, Greece http://www.academy.edu.gr/Antikythera-Digital-Book-Files/0A_The%20Antikythera%20Mechanism_S.pdf, on pages 259–280.

van den Bergh, G. (1955) *Periodicity and Variation of Solar (and Lunar) Eclipses*. Netherlands, Tjeenk Willink and Haarlem.

Bowen, A. C. and Goldstein, B. R. (1988) Meton of Athens and Astronomy in the Late Fifth Century B.C. In: Leichty E., de J. Ellis M., Gerardi P. (eds.), *A Scientific Humanist: Studies in Memory of Abraham Sachs* (Philadelphia), pp. 39–81.

Carman, C. C. and Evans J. (2014) On the epoch of the Antikythera mechanism and its eclipse predictor. *Arch. Hist. Exact Sci.*, 68, pp. 693–774

Carman, C. C. and Di Cocco, M. (2016) The Moon Phase Anomaly in the Antikythera Mechanism, *ISAW Papers*, 11, http://dlib.nyu.edu/awdl/isaw/isaw-papers/11/

EFLIGHT 2021-SUNRISE: A Unique Horizon-Venue "Sunrise" Totality, 2021. http://nicmosis.as.arizona.edu:8000/ECLIPSE_WEB/TSE2021/TSE2021WEB/EFLIGHT2021.html

Edmunds, M. G. (2011) An Initial Assessment of the Accuracy of the Gear Trains in the Antikythera Mechanism, *Journal for the History of Astronomy*, 42, pp. 307–20.

Efstathiou, K., Basiakoulis, A., Efstathiou, M., Anastasiou, M. and Seiradakis J. H. (2011) The Equation of Time calculated by the Antikythera Mechanism. Oral presentation, *4th International Conference on Manufacturing Engineering "ICMEN"*, Thessaloniki, Greece, 3-5 October 2011. Engineers Edge, Measuring Circular Runout, https://www.youtube.com/watch?v=mqqda5FFB9o&t=4s

Espenak, F. and Meeus J. (2008) Five Millennium Catalog of Solar Eclipses: -1999 to +3000 (2000 BCE to 3000 CE), *NASA Tech. Pub.* 2008-214170, NASA Goddard Space Flight Center, Greenbelt, Maryland. NASA Eclipse catalogue,

https://eclipse.gsfc.nasa.gov/SEcat5/SE1901-2000.html,

https://eclipse.gsfc.nasa.gov/SEsearch/SEsearchmap.php?Ecl=-05840528

Fedorov, A., Beichel, R., Kalpathy-Cramer, J., Finet, J., Fillion-Robin, J. C., Pujol, S., Bauer, C., Jennings, D., Fennessy, F., Sonka, M., Buatti, J., Aylward, S. R., Miller, J. V., Pieper S. and Kikinis R. (2012) 3D Slicer as an Image Computing Platform for the Quantitative Imaging Network. *Magn Reson Imaging* 30(9), pp. 1323–1341. PMID: 22770690, https://www.slicer.org/.

Freeth, T. (2019) Revising the eclipse prediction scheme in the Antikythera mechanism. *Palgrave Communications*, 5(7), pp. 1–12.

Freeth, T., Bitsakis, Y., Moussas, X., Seiradakis, J. H., Tselikas, A., Mangou, H., Zafeiropolou, M., Hadland, R., Bate, D., Ramsey, A., Allen, M., Crawley, A., Hockley, P., Malzbender, T., Gelb, D., Ambrisco, W. and Edmunds, M. G. (2006) Decoding the Ancient Greek Astronomical Calculator Known as the Antikythera Mechanism, *Nature*, 444, pp. 587–91.

Freeth, T., Jones, A., Steele, J. M. and Bitsakis Y. (2008) Calendars with Olympiad Display and Eclipse Prediction on the Antikythera Mechanism. *Nature*, 454, pp. 614–7 (Supplementary Material).

Freeth, T. and Jones, A. (2012) The Cosmos in the Antikythera Mechanism, *ISAW Papers*, 11, http://dlib.nyu.edu/awdl/isaw/isaw-papers/4/

Freeth, T., Higgon, D., Dacanalis, A., MacDonald, L., Georgakopoulou, M. and Wojcik, A. (2021) A Model of the Cosmos in the ancient Greek Antikythera Mechanism. *Sci Rep* 11, 5821.

*ASSEMBLING THE FRAGMENT D ON THE ANTIKYTHERA MECHANISM* 127Gourtsoyannis, E. (2010) Hipparchus vs. Ptolemy and the Antikythera Mechanism: Pin–Slot Device Models Lunar Motions, *Advances in Space Research*, 46, pp. 540–4.
Hannah, R. (2001) The Moon, the Sun and the Stars: Counting the Days and the Years. In: McCready S. (ed.), The Discovery of Time. London, MQ Publications, pp. 56–99.
Hannah, R. (2013) Greek Government and the Organization of Time. In Beck H. (ed.), Companion to Ancient Greek Government, Oxford, Blackwell, pp. 349–65.
Iversen, P. and Jones, A., (2019) The Back Plate Inscription and eclipse scheme of the Antikythera Mechanism revisited. *Archive for History of Exact Sciences*, 73, pp. 469–511.
Jones, A. (2017) *A Portable Cosmos*. New York, Oxford University Press.
Total Solar Eclipse of 2021 December 4 in Antarctica, 2021. http://xjubier.free.fr/en/site_pages/solar_eclipses/TSE_2021_GoogleMapFull.html
Kircher, A. (1646) *Ars Magna Lucis et Umbrae in Decem Libros Digesta*, Romæ: Sumptibus Hermanni Scheus; Ex typographia Ludouici Grignani. http://lhldigital.lindahall.org/cdm/ref/collection/color/id/23013
Lazos, C. (1994) *The Antikythera Computer*. Athens, Aiolos Publications.
Manitius, K. (1880) *Gemini Elementa Astronomiae*. Leipzig, in Greek and Latin.
Meeus, J. (1998) *Astronomical Algorithms*. 2nd Edition, Virginia Willmann-Bell,.
Meeus, J. (2004) *Mathematical Astronomy Morsels III*. Virginia, Willmann-Bell.
Meeus, J., Grosjean, C. C. and Vanderleen, W. (1966) *Canon of Solar Eclipses*. United Kingdom, Pergamon Press, Oxford.
Neugebauer, O. (1975) *A History of Ancient Mathematical Astronomy*. Berlin; NewYork, Springer-Verlag.
von Oppolzer, T. R., (1962) *Canon der Finsternisse*. New York, Dover Publications (reprint from 1887 edition).
Panchenko, D. (1994). Thales's Prediction of a Solar Eclipse. *Journal for the History of Astronomy*, 25(4), pp. 275–288.
Pedersen, O. (2011) *A Survey of the Almagest, Sources and Studies in the History of Mathematics and Physical Sciences*. London; New York; Dordrecht; Heidelberg, Springer.
Price, D.S. (1974) Gears from the Greeks: The Antikythera Mechanism, a Calendar Computer from ca. 80 B.C. *Trans. Am. Phil. Soc.* 64(7), pp. 1–70.
Rehm, A. (1905–1906) Notizbuch (unpublished notebooks), research manuscripts and photographs. Bayerische Staatsbibliothek, Munich, Germany. Rehmiana III/7 and III/9.
Roumeliotis, M. (2018) Calculating the torque on the shafts of the Antikythera Mechanism to determine the location of the driving gear. *Mechanism and Machine Theory*, 122, pp. 148-159.
Seiradakis, J. H. and Edmunds, M. (2018) Our current knowledge of the Antikythera Mechanism. *Nature Astronomy* 2(1), pp. 35–42
Spandagos, E. (2002) *Introduction to the Phenomena of Geminus*. Aithra, Athens.
Steele, J. M. (2000a) Eclipse Prediction in Mesopotamia, *Archive for History of Exact Sciences*, 54, pp. 421–54.
Steele, J. M. (2000b) A Re-analysis of the Eclipse Observations in Ptolemy's Almagest. *Centaurus*, 42, pp. 89–108.
Steele, J. M. (2002) A Simple Function for the Length of the Saros in Babylonian Astronomy. In: Steele J.M. and Imhausen A. (eds), Under One Sky: Astronomy and Mathematics in the Ancient Near East. *Ugarit-Verlag*, Münster, pp. 405–420.
Steele, J. M. (2015) Eclipses: Calculating and Predicting Eclipses. In: Selin H. (eds), Encyclopaedia of the History of Science, Technology, and Medicine in Non-Western Cultures. *Springer*, Dordrecht.
Stephenson, R. F. and Fatoohi L. J. (1997) Thales's Prediction of a Solar Eclipse, *Journal for the History of Astronomy*, 28(4), pp. 279–282.
The Great American Eclipse, https://www.greatamericaneclipse.com/basics
Theodosiou, S. and Danezis, M. (1995) *The Calendar Odyssey*, in Greek. Athens, Diaylos.
Toomer, G. J. (1984) *Ptolemy's Almagest. Edition. and translation*. London: Duckwort.
Vaughan, V. (2002) *The Origin of the Olympics: Ancient Calendars and the Race Against Time*. Massachusetts, One Reed Publications.
Voulgaris, A., Vossinakis, A. and Mouratidis, C. (2018a) The New Findings from the Antikythera Mechanism Front Plate Astronomical Dial and its Reconstruction. *Archeomatica International*, Special Issue 3(8), pp. 6–18. https://www.yumpu.com/en/document/view/59846561/archeomatica-international-2017.
Voulgaris, A., Mouratidis, C. and Vossinakis, A. (2018b) Conclusions from the Functional Reconstruction of the Antikythera Mechanism. *Journal for the History of Astronomy*, 49(2), pp. 216–238.Mediterranean Archaeology and Archaeometry, Vol. 22, No 3, (2022), pp. 103-131

# SUPPLEMENTARY MATERIAL

We present a further analysis regarding the position of Fragment D parts, especially for the position of the perpendicular stabilizing pin. According to the AMRP tomographies the stabilizing pin (today is displaced) is located between the gear r1 and the Additional plate. In order to better clarified the position of the parts, we used bronze reconctructions of the parts, adapted on their original placement. An analysis of the parts position according to Freeth et al., (2021) suggested design of the Fragment D is also presented and discussed and concluding remarks are drawn.

1. **Technical information and data comparison regarding the parts position of Fragment D, based on AMRP CT's in Figs. 1, II.**

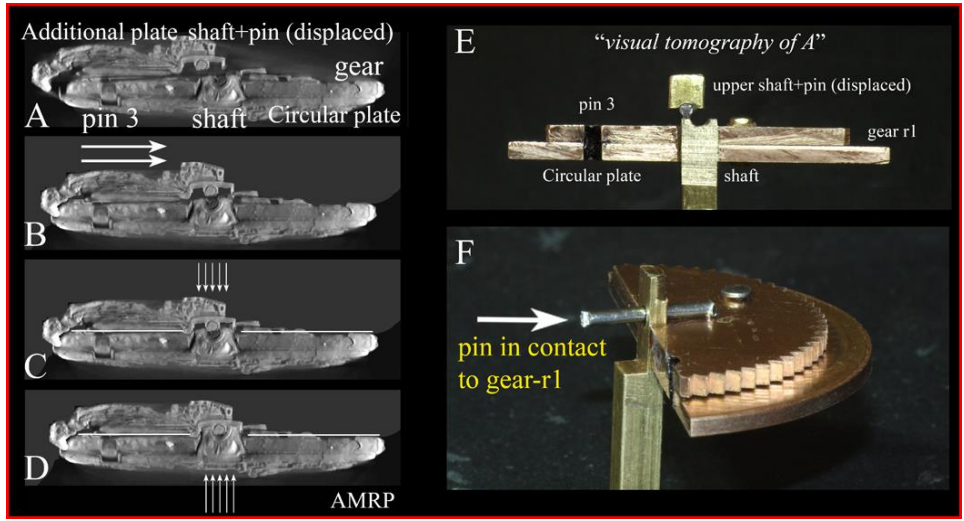

*Figure I. A,B,C,D, The reposition of the broken upper part of the r-shaft. The horizontal white lines define the upper surface of r-gear. The pin is in contact to gear-r1 (before its displaced position). E) A "Visual tomography" using bronze parts, represnting the Computed Tomography of A (the parts of Fragment D, were constructed in bronze and afterwards were cut in half). F) After the restoration of the broken/displaced upper part, the initial position of the pin is represented in bronze parts. The pin is in contact with the upper surface of r-gear. Bronze parts construction and images by the first author.*

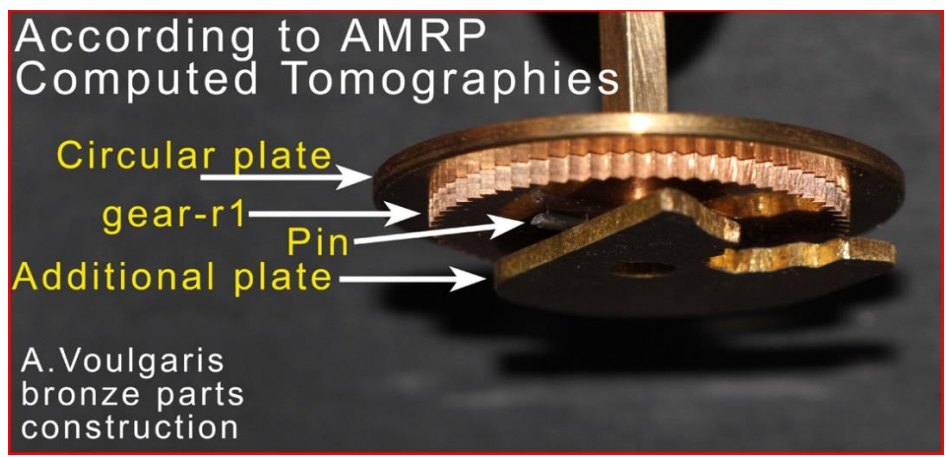

*Figure II. Fragment D reconstruction in bronze, based on AMRP CTs (here is presented in inverted position).*

*Parts position from top to bottom:*

1. *Circular plate*
2. *Gear-r1*
3. *Stabilizing Pin, perpendicular to r-axis*
4. *Additional plate*

*Remarks: A gap (in the thickness of pin) exists between gear r1 and Additional plate (they are not in contact).*

## 2. Freeth et al., (2021) suggestion scheme for Fragment D, in Figs. III, IV.

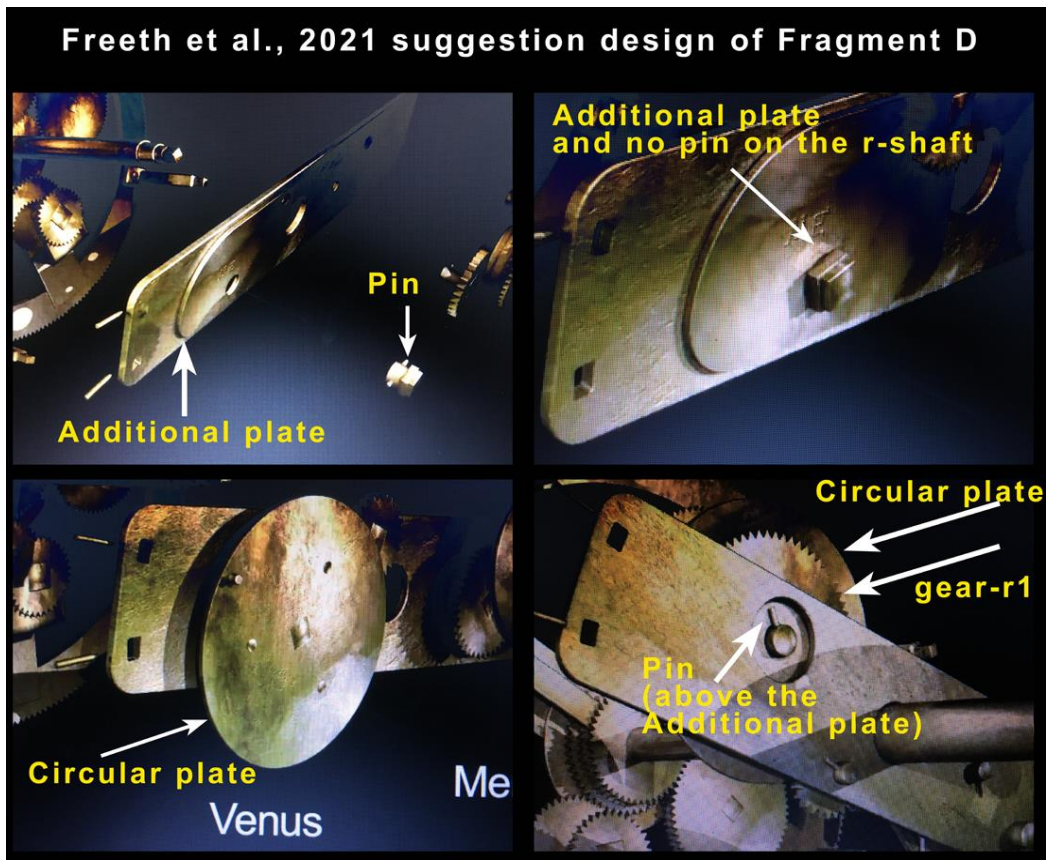

*Figure III. Freeth et al., (2021) 3D representation of Fragment D, suggested for Venus output. The perpendicular stabilizing pin is presented as the last part of the Fragment D (after the Additional plate). Data from Freeth et al. (2021) Supplementary Information 1, Venus output gearing video frames @ 0:28 – 0:34).*

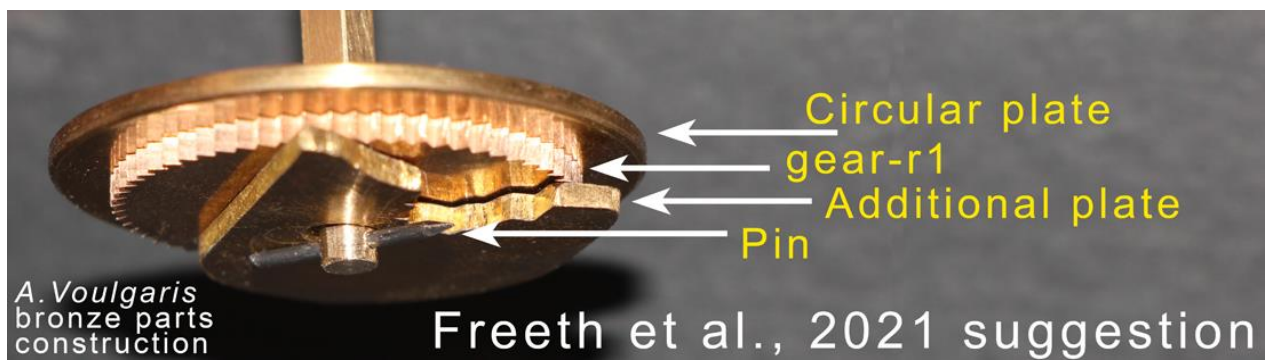

*Figure IV. First author bronze reconstruction based on Freeth et al., (2021) 3D representation (see previous Fig. III).*

*Parts position from top to bottom:*

1. *Circular plate*
2. *Gear-r1*
3. *Additional plate*
4. *Stabilizing Pin perpendicular to r-axis*

*Remarks: in Freeth et al., (2021), the gear r1 is in contact to the Additional plate. There is no exists a gap.*

Freeth et al., (2021) present the Additional plate in contact to gear-r1 (Figure 4p of Freeth et al., 2021 and in Figure S14 of their *Supplementary Information*, also in their *Supplementary Information 1*, video at 0:28 – 0:34) which is not in agreement to AMRP tomographies of Fragment D (see Fig. I and II). In Author's work is presented a further discussion regarding the functionality of the Fragment D in inverted position (as presented here) in *Section 5.2 A further analysis and discussion for Fragment D parts*.

## 3. Fragment D design in authors present work *vs* Freeth et al., (2021), in Fig.V.

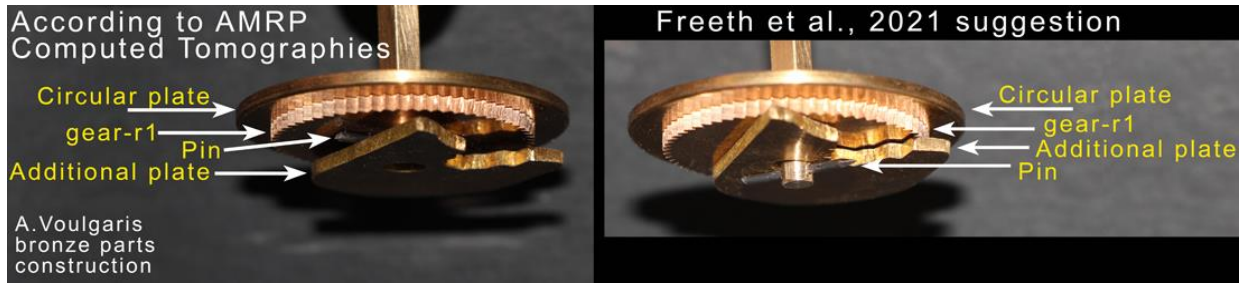

*Figure V. Left: the position of perpendicular stabilizing pin according to AMRP CT's of Fragment D (between gear r1 and the Additional plate). Right: the position of pin in Freeth et al., (2021) 3D suggested representation (after the Additional plate).*

## 4. Concluding Remarks

1. The position of the pin as the last part of Fragment D (as suggested by Freeth et al., 2021) is not in agreement according to AMRP CTs.
2. Generally: Changing the position of a mechanical part, the mechanical status, the mechanical properties and the functionality of the system, change.
   By changing the position of the pin (i.e. the mechanical design) the mechanical status of the construction changes, leading to a different mechanical behavior, results and observations, which could be deviated by the original.
3. By moving the pin to its correct position (i.e. between the gear r1 and the Additional plate), strong mechanical problems about the stability and the functionality of inverted Fragment D (the Seesaw effect, see Fig. 12), as the gearing of Venus in the Antikythera Mechanism, appear (see *Section 5.2 A further analysis and discussion for Fragment D parts*).
4. The invert position of Fragment D, as suggested by Freeth et al., (2021) (after the pin adaptation on its correct position) is doubtful, appears mechanical problems, it is unorthodox in mechanical terms and does not agree with the constructional characteristics inferred by the preserved gears.

## Data

Antikythera Mechanism X-Ray Data was provided to the authors by J.H. Seiradakis and X. Moussas, members of Antikythera Mechanism Research Project.
Video data retracted by Freeth et al., (2021) *Supplementary Information 1, Venus output gearing, video frames @ 0:28 – 0:34)* https://www.nature.com/articles/s41598-021-84310-w.